\documentclass[10pt]{iopart}
\pdfoutput=1

\usepackage[utf8]{inputenc}
\usepackage{graphicx}
\usepackage{xcolor}
\usepackage{booktabs}
\usepackage{subcaption}
\usepackage{setspace}
\ioptwocol
\usepackage{multirow}

\usepackage{lipsum}

\begin{document}

\title[Dynamics of OCT4 in hESCs]{OCT4 expression in human embryonic stem cells: spatio-temporal dynamics and fate transitions}

\author{L E Wadkin$^{\#, 1}$, S Orozco-Fuentes$^{\#,1,2}$, I Neganova$^3$, M Lako$^4$, R A Barrio$^5$, A W Baggaley$^1$, N G Parker$^1$ and A Shukurov$^1$}
\address{$^1$ School of Mathematics, Statistics and Physics, Newcastle University, Newcastle upon Tyne, UK.}
\address{$^2$  Department of Mathematics, Physics and Electrical Engineering, Northumbria University, Newcastle upon Tyne, UK.}
\address{$^3$ Institute of Cytology, RAS St Petersburg, Russia.}
\address{$^4$ Biosciences Institute, Faculty of Medical Sciences, Newcastle University, Newcastle upon Tyne, UK.}
\address{$^5$ Instituto de Física, Departamento de Física-Química, Universidad Nacional Autónoma de México, Mexico City, Mexico.}
\address{$\#$ Contributed equally}
\ead{l.e.wadkin@ncl.ac.uk, sirio.orozco-fuentes@northumbria.ac.uk}
\vspace{10pt}
\begin{indented}
	\item[]October 2020
\end{indented}

\begin{abstract}
	The improved \textit{in-vitro} regulation of human embryonic stem cell (hESC) pluripotency and differentiation trajectories is required for their promising clinical applications. The temporal and spatial quantification of the molecular interactions controlling pluripotency is also necessary for the development of successful mathematical and computational models.
	
	Here we use time-lapse experimental data of OCT4-mCherry fluorescence intensity to quantify the temporal and spatial dynamics of the pluripotency transcription factor OCT4 in a growing hESC colony in the presence and absence of BMP4. We characterise the internal self-regulation of OCT4 using the Hurst exponent and autocorrelation analysis, quantify the intra-cellular fluctuations and consider the diffusive nature of OCT4 evolution for individual cells and pairs of their descendants. We find that OCT4 abundance in the daughter cells fluctuates sub-diffusively, showing anti-persistent self-regulation.
	
	We obtain the stationary probability distributions governing hESC transitions amongst the different cell states and establish the times at which pro-fate cells (which later give rise to pluripotent or differentiated cells) cluster in the colony. By quantifying the similarities between the OCT4 expression amongst neighbouring cells, we show that hESCs express similar OCT4 to cells within their local neighbourhood within the first two days of the experiment and before BMP4 treatment. 
	
	Our framework allows us to quantify the relevant properties of proliferating hESC colonies and the procedure is widely applicable to other transcription factors and cell populations.

\end{abstract}


\submitto{\PB}

\section{Introduction}

Human pluripotent stem cells (hPSCs), encompassing human embryonic stem cells (hESCs) and human-induced pluripotent stem cells (hiPSCs) self-renew indefinitely while maintaining the property to give rise, under differentiation conditions, to almost any cell type in the human body \cite{TAKAHASHI2007861,Li04,DANIELS20103563}.  The maintenance and control of the pluripotency and differentiation trajectories of hPSCs is central to their touted applications in drug discovery, and regenerative and personalised medicine \cite{Bauwens08,Ebert10,Zhu13,Avior16,Ilic17,Shroff17,Trounson16}. 

The control and optimisation of pluripotency across colonies is difficult due to its complex inter-regulatory dynamics. This regulatory network consists of a core set of pluripotency transciption factors (TFs) expressed to maintain self-renewal and suppress differentiation \cite{Young05,Chambers09}. Amongst the most important TFs that preserve the undifferentiated state in hESCs are NANOG, OCT4 and SOX2 \cite{Li04,Young05,Chambers09}. During development, these TFs become expressed at different levels, initiating differentiation towards specific cell lineages following signalling cues \cite{Pauklin13}. Pluripotency is also affected by external factors: the local environment \cite{Shuzui19,Stadhouders19}, interactions with neighbours \cite{Nemashkalo17,Rosowski15}, the cell cycle \cite{Pauklin13} and the substrate \cite{Hwang08}. On the colony scale, complex collective effects of pluripotency emerge. In the presence of restrictive geometries, differentiated cells form bands occurring around colony edges \cite{Rosowski15,Warmflash14}.

Several experiments have been performed to quantify the behaviour and joint influence of each TF in the pluripotent cell \cite{ZhanYuOCT4, RodriguezRegulationOCT,Guilai2010,GuangjinThomson}. Their results indicate that the expression of the TF proteins are highly variable both at the single-cell (time) and colony-level (space) due to intrinsic noise in gene expression, interactions at the molecular level and the randomness present in the extracellular environment \cite{Lin18, Wu12,LAURENT2011106}. Thus, heterogeneity and stochasticity are inherent properties of pluripotent stem cell populations \cite{Wu12,Chambers14, HuangDev} that hinder their clonal expansion in culture \cite{Stewart2006,Dakhore2018,Stockholm07,Li12}.

OCT4, in conjunction with other core members of the pluripotent regulatory network, activates both protein-coding genes and non-coding RNAs necessary to maintain pluripotency \cite{Guilai2010}. In mouse embryonic stem cells (mESCs), OCT4 expression is relatively uniform with a high correlation between its levels and pluripotency \cite{Navarro12,Munoz12}. In hESCs OCT4 also interacts with the BMP4 (bone-morphogenetic protein) pathway. Under standard culture conditions BMP4 acts as a morphogen \cite{Nemashkalo17} and defines several cell fates: in the presence of BMP4, high levels of OCT4 promote mesendoderm differentiation, while low levels result in extra-embryonic ectoderm and primitive endoderm differentiation \cite{Guilai2010,Wang12,Radzisheuskaya2014}.

Fluorescently-labelled hESCs are useful tools for the {\it in-vitro} tracking, visualisation and real-time monitoring of the hESCs without the need for cellular fixation. These single-cell measurements can be used for accurate quantification of the protein changes in time and space. Recent studies of the expression of OCT4 in hESCs bearing the OCT4‐mCherry reporter \cite{Ovchinnikov} indicate inheritance of similar protein levels from mother to daughter cells, with the OCT4 levels established in new daughter cells being predictive of long-term cellular states \cite{Wolff18}. Although the daughter cells continue to be very similar to their predecessors, in the long term, further variations get amplified with consecutive cell divisions and thus the heterogeneous hESC population is established by incremental divergences \cite{Setty}. These divergences, caused by regulatory mechanisms, noise in the protein expression, etc., create paths through all possible cell states (fates) which result in distinctive patterns in hESC colonies under differentiating conditions \cite{Rosowski15,Tewary4298}.

These observations raise questions about the temporal behaviour of the OCT4 signal and how and when the cell fates get established within a hESC colony. In this paper we build upon the previously published work of Ref. \cite{Wolff18} which considers time-lapse fluorescent measurements of the OCT4-mCherry reporter levels in cells in a growing hESC colony. Although the dynamics of OCT4 are complex, affected by many genetic factors and closely regulated by the other TFs \cite{Li04,Wang12,Babaie07}, here we isolate autonomous properties of OCT4 to facilitate the development of descriptive mathematical models.

We describe quantitatively the fluctuations in OCT4 in relation to cell fate and the addition of the differentiation agent BMP4. We quantify the self-regulation of OCT4 through anti-persistence and characterise it within the diffusion framework. Using custom-designed software, we reconstruct the hESC colony spatio-temporally, examine the establishment of the hESCs pro-fates (pluripotent, differentiated and unknown) and report the transition probability matrices of the hESCs between the different pro-fates at mitosis. These matrices result in the stationary distributions of cell fates that get established in the hESC colony in the presence and absence of BMP4.

From our spatial analyses of the hESCs positions within the colony, we calculate the time at which the cells segregate in terms of their pro-fates. This gives a time-frame for the emergence of pre-patterning in a hESC colony. Finally, we quantify the `cooperation' between nearest hESCs, defined in terms of a dissimilarity metric between their OCT4 values. Our quantitative analyses, along with Ref. \cite{Wolff18} provides a transferable basis for the comparison to other TFs and for developments in mathematical and statistical models of pluripotency.

\section{Materials and Methods}

\subsection{Experiment}
\begin{figure*}[h]
	\begin{center}
		\includegraphics[width=\textwidth]{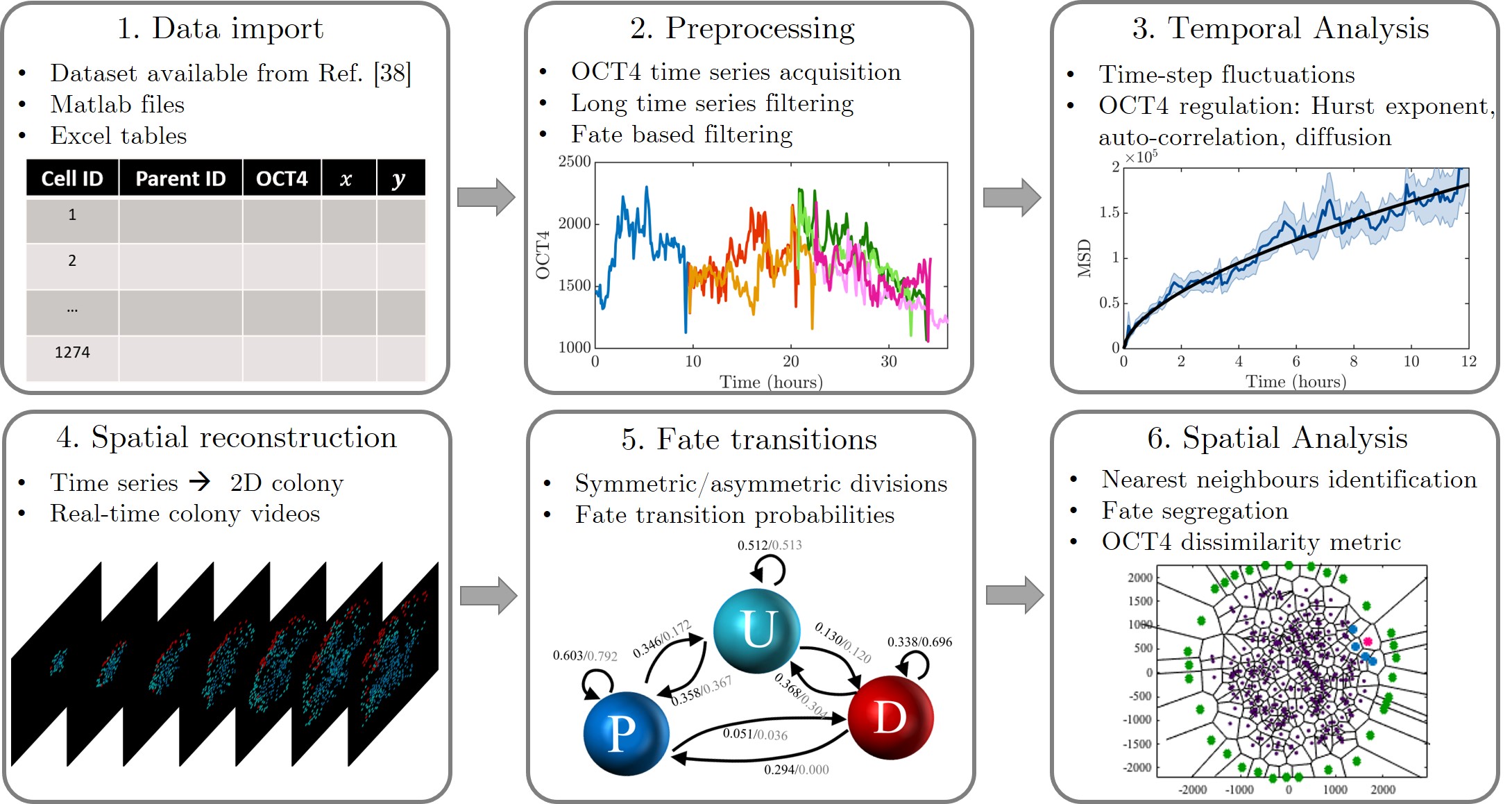}
		\caption{Workflow diagram showing the steps taken to analyse the experimental dataset from Ref. \cite{Wolff18}.}
		\label{fig:workflow} 
	\end{center}
\end{figure*}

The experiment was carried out by Purvis Lab (University of North Carolina, School of Medicine), and published in Ref. \cite{Wolff18}. The experiment details are described thoroughly in Ref. \cite{Wolff18}. In the following, we give a brief description to facilitate the reading of our paper. We show a workflow diagram to illustrate our steps in analysing the experimental data set in Figure~\ref{fig:workflow}.

The OCT4 levels (time-lapse mean OCT4-mCherry fluorescence intensity) in a hESC (H9) colony were determined over multiple generations until their differentiation to extra-embryonic mesoderm. Cells were live-imaged for 68 hours. The experiment begins at $t_\mathrm{exp} = 0\,$h and ends at $t_\mathrm{exp}= 68\,$h. At $t_\mathrm{exp} = 43\,$h, the hESCs are treated with (100\,ng/ml) bone-morphogenetic protein 4 (BMP4) to induce their differentiation towards distinct cell fates. Therefore, time $t_\mathrm{exp}<43\,$h indicates the absence, and $t_\mathrm{exp}>43\,$h the presence of BMP4.

The data is provided as time series and includes the position ($x(t)$, $y(t)$), radial position within the colony ($r(t)$) and OCT4 immunofluorescence intensity values ($\Omega(t)$) of each cell at intervals of $\Delta t = {5}\,$min. The initial and final times in the time series denote the cell birth and division, respectively. The OCT4 values are reported in arbitrary fluorescence units (a.f.u.), the intensity values in terms of the number of photons detected by the microscope from the specimen.

To classify the cells as either self-renewing (pluripotent) or differentiated, the expression levels of CDX2 were quantified at $t_\mathrm{exp} = 68\,$h. The CDX2 levels, along with the final OCT4 expressions were used to classify the cells according to their pro-fates using a two‐component mixed Gaussian distribution, that is, those cells belonging exclusively to a pluripotent (self-renewing) or differentiated state. A remaining group of uncatalogued cells were classified in an unknown category.  Using these pro-fates, the cell population was traced back in time, spanning multiple cell divisions, with each earlier cell labelled according to this pro-fate. We use the letters P, D and U to denote the pluripotent, differentiated and unknown pro-fates, respectively.

\subsection{Quantitative analysis}
The quantitative analysis in both Ref. \cite{Wolff18} and this manuscript were performed using MATLAB\textsuperscript{\textregistered}. We give full details of the quantitative methods in the Supplementary Information.\\ 

 \section{Results}

\subsection{Colony summary}

The colony begins from 30 cells and grows over 68\,hours (817 time frames) to 463 cells, with 1274 cell cycles elapsing within this time. The number of cells, $N$, considered in each cell pro-fate category, pre- ($t_\mathrm{exp}<43\,$h) and post-BMP4 ($t_\mathrm{exp}>43\,$h) addition is given in Table~\ref{tab:Ncells}. An analysis of the number of cells in the colony over time, $N(t_\mathrm{exp})$, is given in the Supplementary Information (Figure S1). The whole colony follows exponential growth, with a doubling time of $16\pm0.01$ hours, as noted in Ref. \cite{Wolff18} and consistent with other reports \cite{Ghule11,Wadkin19}. When split by fate, the pluripotent cells proliferate significantly faster than the differentiated cells.

\begin{table}[h]
	\centering
		\begin{tabular}[t]{cccc}
			\toprule
			$N$ &Pre-BMP4&Post-BMP4&All times\\
			\midrule
			P&96&422&518\\
			D&22&111&133\\
			U&112&511&623\\
			All fates& 230 & 1044 & 1274\\
		\bottomrule
	\end{tabular}
	
	% Or to place a caption below a table
	\caption{The number of cells, $N$, in each of the cell fate (pluripotent P, differentiated D and unknown U) and pre- and post-BMP4 categories. A post-BMP4 is any cell present at $t_\mathrm{exp}=43\,$h or later. There are 1274 cell cycles in total.}
	\label{tab:Ncells}
\end{table}%

\begin{figure*}[!]
	\centering
	\begin{subfigure}[t]{1\textwidth}
		\includegraphics[width=\textwidth]{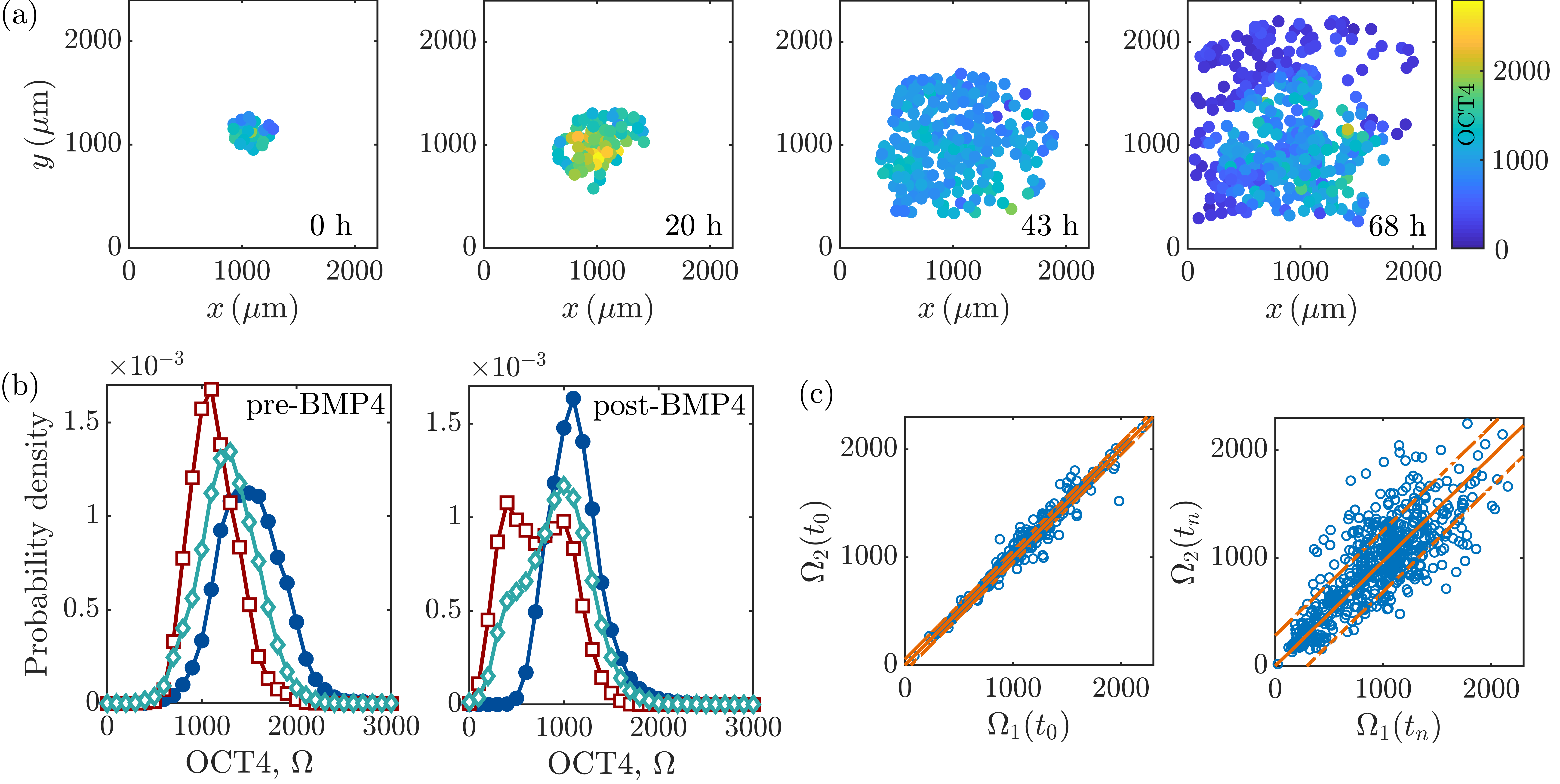}
	\end{subfigure}	
	\caption{\label{fig:col}(a) Snapshots of the colony at $t_\mathrm{exp}=0\,$h, $20\,$h, $43\,$h (addition of BMP4) and $68\,$h (final time). The cells are coloured according to their OCT4 intensity levels. Note that the circles are not indicative of cell or nucleus size. (b) The distributions of OCT4 for pluripotent (filled blue circles), differentiated (open red squares) and unknown cells (open turquoise diamonds), pre- and post-BMP4. (c) OCT4 values for all sister pairs (564 pairs) at the start ($\Omega_i(t_0)$) and end ($\Omega_i(t_n)$) of their cell cycles. The lines of best fit (orange solid lines) with standard errors in predicting a future observation (dashed lines) are $\Omega_1(t_0)=(1\pm0.003)\Omega_2(t_0)$ with $R^2=0.98$ and $\Omega_1(t_n)=(0.97\pm0.2)\Omega_2(t_n)$ with $R^2=0.78$ for initial and final OCT4, respectively.}
\end{figure*}

Spatial reconstruction allows us to visualise the evolution of the colony as in Video S1 (cells labelled by pro-fate) and Video S2 (cells labelled by OCT4). Corresponding snapshot images at the beginning of treatment with BMP4 and at the end of the experiment are shown in the Supplementary Information, Figure S2. Snapshots of the colony colour-coded by OCT4 intensity are shown in Figure~\ref{fig:col}(a). Spatial clustering of pro-fate and OCT4 can be seen.

The measurement of the OCT4 signal at 5 minute intervals, results in a set of evenly sampled discrete observations for each cell, $\Omega(t_0), \Omega(t_1), ..., \Omega(t_n)$, where $t_0$ and $t_n$ denote the times of cell birth and division, respectively. The values of $t_n$ range from 15 minutes to 30 hours across the population. The OCT4 time series for a cell at the beginning of the experiment and its descendants is shown in Figure~\ref{fig:workflow}. 

The distributions of all measured OCT4 values are shown in Figure~\ref{fig:col}(b). The mean, median and kurtosis of each distribution are given in Table~\ref{tab:OCT}. Pro-differentiated cells pre-BMP4 show a visibly skewed distribution, with a higher preference of lower OCT4 expressions than the pluripotent cells, fitting with the pre-determined fate choice identified in Ref. \cite{Wolff18}. Post-BMP4, all fates also show a reduction in their OCT4 expression, with the effect seen most strongly in the differentiated cells. This reduction in OCT4 in the differentiated cells is expected, but it is interesting that the same effect (to a lesser extent) is also present in the other fate groups.

\begin{table}[h]
	\centering
	\resizebox{\columnwidth}{!}{%
		\begin{tabular}[t]{cccc}
			\toprule
			& & Pre-BMP4&Post-BMP4\\
			\midrule
			\multirow{3}{*}{P} & Mean & 1510 $\pm$ 230 & 1120 $\pm$ 270 \\
			& Median & 1500 [1280 1730] & 1090 [930 1260] \\
			& Kurtosis & 3.3 & 5.2 \\
			\midrule
			\multirow{3}{*}{D} & Mean & 1130 $\pm$ 240 & 730 $\pm$ 320 \\
			& Median & 1100 [960 1290]& 720 [450 990] \\
			& Kurtosis & 3.3 & 2.1 \\
			\midrule
			\multirow{3}{*}{U} & Mean & 1280 $\pm$ 300 & 910 $\pm$ 350\\
			& Median & 1280 [1090 1490] & 940 [670 1150]\\
			& Kurtosis & 3.0 & 3.1 \\
			\bottomrule
		\end{tabular}
	}
	\caption{The mean $\pm$ standard deviation, median [lower quartile upper quartile] and kurtosis of the OCT4 distributions for each pro-fate (pluripotent P, differentiated D and unknown U) shown in Figure~\ref{fig:col}(b).}
	\label{tab:OCT}
\end{table}%   

The corresponding temporal probability density functions for OCT4 over time are given in the Supplementary Information, Figure S3. We also present the average OCT4 expression with time at the colony level in Figure S4. Both show the reduction of OCT4 expression in all pro-fates with time.

The analysis in Ref. \cite{Wolff18} shows that upon cell division the OCT4 ratio between sister cells is centred around 1:1, meaning that although asymmetric inheritance does occur, on average sister cells start with similar levels of OCT4. We consider the correlation in the OCT4 time series for sister cells over their lifetimes by calculating the correlation coefficient, $\rho$. Note that for sister pairs with unequal cell-cycle times, the correlation is calculated for the time series of the length up to the minimum cell cycle time. Each OCT4 time series was de-trended to account for any confounding similarities in sister cells that may be present due to their shared environment. The distributions of $\rho$ are shown in the Supplementary Information, Figure S5. The mean correlations, $\overline{\rho}$, are given in Table~\ref{tab:corr} and show moderate positive correlations across all categories.

All pro-fates show $\overline{\rho}=0.5$ ($\pm0.2$, $\pm0.3$ and $\pm0.3$ for pluripotent, differentiated and unknown cells, respectively).  Sister cells pre-BMP4 show a weaker correlation than those post-BMP4, with $\overline{\rho}=0.3\pm0.2$ and $\overline{\rho}=0.5\pm0.3$, respectively. This suggests that BMP4 treatment exacerbates the similarities in sister cell OCT4. These results quantify the regulation between sister cells and further illustrate that this regulation is systematic and importantly, still present when confounding external trends are removed. 

\begin{table}[h]
	\centering
	\resizebox{\columnwidth}{!}{%
		\begin{tabular}[t]{cccc}
			\toprule
			$\overline{\rho}$&Pre-BMP4&Post-BMP4& All times\\
			\midrule
			P&-&-&$0.5\pm0.3(0.02)$\\
			D&-&-&$0.5\pm0.2(0.04)$\\
			U&-&-&$0.5\pm0.3(0.02)$\\
			All& $0.3\pm0.2(0.03)$ & $0.5\pm0.3(0.01)$ & $0.5\pm0.3(0.01)$\\
			\bottomrule
		\end{tabular}
	}
	% Or to place a caption below a table
	\caption{The mean correlation, $\overline{\rho}$ $\pm$ the standard deviation (standard error) between pairs of sister cells.}
	\label{tab:corr}
\end{table}%

We can also quantify how this correlation between sister cells varies through their lifetimes. The initial ($\Omega_i(t_0)$) and final ($\Omega_i(t_n)$) OCT4 values for all sister cells are shown in Figure~\ref{fig:col}(c). The initial values follow a close relationship (consistent with the OCT4 ratio splitting distribution in Ref. \cite{Wolff18}), with correlation $\rho=0.99$ and the trend line $\Omega_1(t_0)=(1\pm0.003)\Omega_2(t_0)$. Note that the labelling of the sister cells as cell 1 and cell 2 is arbitrary. By the end of their respective lifetimes, the distribution spreads, with $\rho=0.78$ and a line of best fit $\Omega_1(t_n)=(0.97\pm0.2)\Omega_2(t_n)$.

Next, we consider the behaviour of OCT4 from the initial point of possible asymmetric inheritance to the final time before mitosis and characterise how this drift of similarity shown in Figure~\ref{fig:col}(c) occurs.

\subsection{Temporal OCT4 dynamics}

In this section we quantify the temporal behaviour of OCT4 dynamics on the cellular level over the course of a cell lifetime. We consider the variability between discrete time-steps and quantify the self-regulatory behaviour of OCT4 using several methods. 

\subsubsection{Variability at short time scales}\hfill

\noindent
Even small fluctuations in TF abundance impact cell fate \cite{Strebinger19}, with both high and low TF values resulting in differentiation \cite{Niwa00,Kopp08}. The mathematical quantification of TF fluctuation will facilitate the description of pluripotency over discrete time-steps, fitting for time-lapse experiments such as the one considered here \cite{Wolff18}. 

\begin{figure*}[!]
	\centering
	\begin{subfigure}[t]{1\textwidth}
		\includegraphics[width=\textwidth]{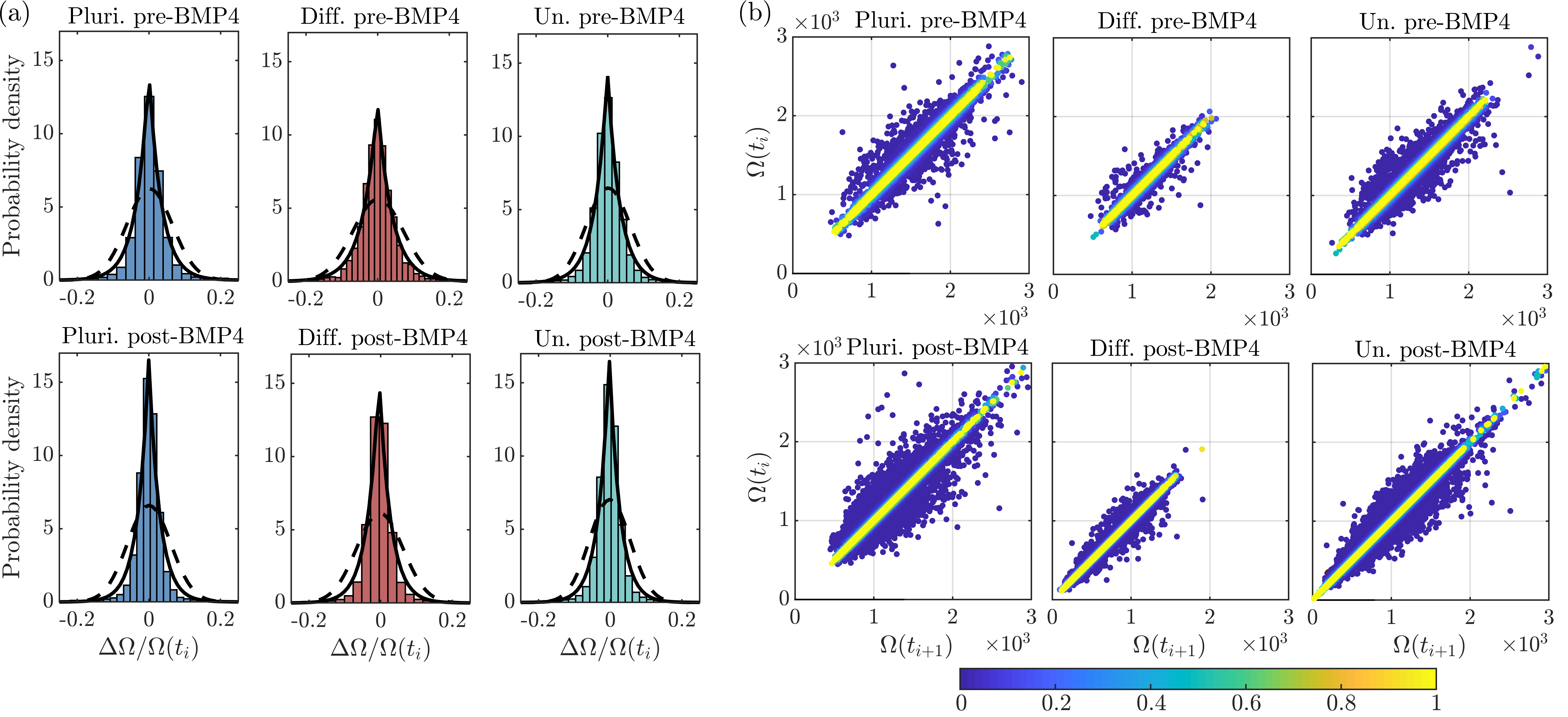}
	\end{subfigure}	
	\caption{\label{fig:deltaoct}(a) Distributions of the change in OCT4 between time frames ($\Delta\Omega/\Omega(t_i)$). Solid lines show the Laplace distribution fittings, given in Table~\ref{tab:laplace} and dashed lines show the Normal distribution fittings. (b) The Poincar\'e maps for the OCT4 signal. The colour bar shows the normalised relative frequency of the points.}
\end{figure*}

We denote the change in the intra-cellular OCT4 abundance between the five minute intervals as $\Delta\Omega=\Omega(t_{i+1})-\Omega(t_{i})$. We consider the scaled distributions, $\Delta\Omega/\Omega(t_i)$, shown in Figure~\ref{fig:deltaoct}(a), to account for the reduction in expression post-BMP4 addition. The scaled fluctuations are centred around zero, however, the distributions are not Normal (confirmed by the Kolmogorov-Smirnov and Shapiro-Wilk tests at the 95\% confidence level) due to a narrower and steeper peak. A Laplace distribution, Laplace$(\mu^\dagger,b)$, better fits the experimental data in all cases, with the parameters given in Table~\ref{tab:laplace}. The distributions of $\Delta\Omega$ are shown in the Supplementary Information, Figure S6 with fittings in Table S1. 

\begin{table}[h]
	\centering
		\resizebox{\columnwidth}{!}{%
	\begin{tabular}[t]{ccc}
		\toprule
		Laplace($\mu^\dagger$,$b$)&Pre-BMP4&Post-BMP4\\
		\midrule
		P& ($-4.6\times10^{-4}$, $0.035$) &($-2.4\times10^{-4}$, $0.030$)\\
		D&($-7.1\times10^{-4}$, $0.041$) & ($-4.6\times10^{-3}$, $0.032$) \\
		U&($-7.8\times10^{-4}$, $0.035$) & ($-2.5\times10^{-3}$, $0.030$)  \\
		\bottomrule
	\end{tabular}
}
	% Or to place a caption below a table
	\caption{The parameters from the Laplace($\mu^\dagger$,$b$) fittings to the $\Delta\Omega/\Omega(t_i)$ distributions shown in Figure~\ref{fig:deltaoct}(a).}
	\label{tab:laplace}
\end{table}%   

Post-BMP4 addition, the distributions for all pro-fates become narrower, with the parameter $b$ showing reductions of 14\%, 22\% and 15\% for pluripotent, differentiated and unknown cells, respectively. This narrowing is due to a preference of smaller changes in OCT4 provoked by the differentiation agent. This could be driven by induced selectivity caused by the BMP4 addition (i.e., the BMP4 causes a systematic change, producing a preference for smaller  $\Delta\Omega/\Omega(t_i)$ values), or it could suggest some collective self-regulation. It is expected, since the differentiated cells are most affected by the BMP4, that this group would show the biggest reduction in variation and therefore the strongest regulation in their OCT4 values. Averaged over all cells, $\overline{\Delta\Omega/\Omega(t_i)}$, (Supplementary Information, Figure S6) shows that although the trend is strongest post-BMP4, there are periods of sustained negative fluctuations from as early as 5 hours.
 
We can also consider the self-similarity of the OCT4 series using the Poincar\'e map \cite{Burykin14, Fishman12}. For each cell, its OCT4 time series can be plotted against itself with one time-step delay, i.e., $\Omega(t_i)$ against $\Omega(t_{i+1})$, shown in Figure~\ref{fig:deltaoct}(b). By assessing qualitatively the shape formed by the return map, we observe changes in the distribution of points pre- and post-BMP4. Even pre-BMP4 addition, the differentiated cells show less variation compared to the pluripotent cells, with the addition of BMP4 exacerbating this effect. Quantitatively this can be described by fitting an ellipse to the shape formed by the data plots, given in the Supplementary Information Table S2.

This information quantifies step changes in OCT4, suggesting Laplace distributions to simulate variation and showing that the addition of BMP4 provokes tighter self-regulation across all cell fates. It also highlights that even between small time increments such as these, the fluctuations post BMP4 should be considered separately for cells of different fates, not only in terms of their average, as expected, but also their variability.

\subsubsection{OCT4 self-regulation}\hfill

\noindent
To investigate the self-regulation and internal memory of OCT4 during a cell cycle, we consider three related approaches, the Hurst exponent, the autocorrelation function and diffusion analysis.

\paragraph{The Hurst exponent}
The Hurst exponent, $0<H<1$ is a measure of the long term memory of a time series \cite{Mandelbrot68,Mielniczuk07}. If a series is Brownian, $H=0.5$, with mutually statistically independent fluctuations. If the series is persistent, $H>0.5$, and at each time-step the series is more likely to fluctuate in the same direction as in the previous step, i.e., if in the last time-step there was an increase, it is more likely there will be another increase during the next time-step. For anti-persistence, $H<0.5$, the series is less likely to fluctuate in the same direction as the previous step. 

\begin{figure}[!]
	\centering
	\begin{subfigure}[t]{0.49\textwidth}
		\includegraphics[width=\textwidth]{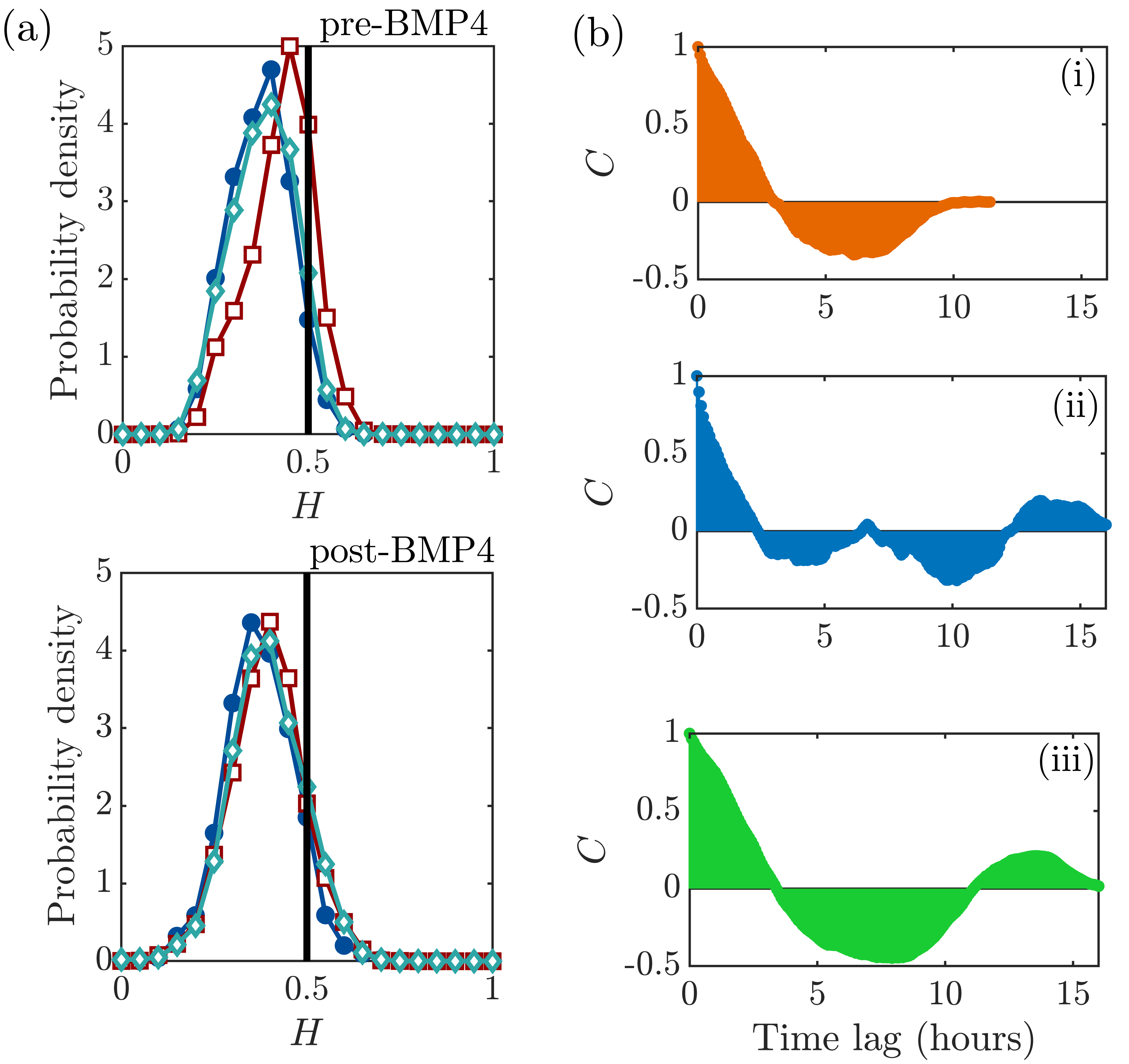}
	\end{subfigure}	
	\caption{\label{fig:regulation}(a) The Hurst exponent, $H$, for pluripotent cells (filled blue circles), differentiated cells (open red squares) and unknown cells (open turquoise diamonds). The black lines show $H=0.5$ for Brownian fluctuations. (b) Typical autocorrelations showing (i) a period of anti-correlation before settling at zero (seen in 51\% of cells, blue solid), (ii) two periods of anti-correlation followed by correlation (seen in 28\% of cells) and (iii) a period of anti-correlation followed by a period of correlation (seen in 14\% of cells).}
\end{figure}

We calculate the Hurst exponent for all cells which live longer than 50 time frames (4.16 hours). The distributions of all $H$ values are shown in Figure~\ref{fig:regulation}(a) with the average Hurst exponents, $\overline{H}$, given in Table~\ref{tab:hurst}. In all cases, the Hurst exponents are significantly less than 0.5, showing moderate anti-persistence indicative of intra-cellular OCT4 self-regulation. Note that this describes the behaviour only on the time scale of a single cell lifetime and that there are other longer-term behaviours also influencing expression \cite{Wolff18}. There is no significant difference in $H$ before and after the BMP4 addition in all cell fates (confirmed by the Kolmogorov-Smirnov test at the 95\% level) suggesting this aspect of the self-regulatory behaviour is inherent to the cells and unchanged by the differentiation stimulus. 

\begin{table}[h]
	\centering
	\begin{tabular}[t]{ccc}
		\toprule
		$\overline{H}$ & Pre-BMP4 & Post-BMP4\\
		\midrule
		P& 0.37 (0.08 0.008)  & 0.37 (0.09 0.004) \\
		D& 0.42 (0.08 0.02)& 0.39 (0.09 0.009) \\
		U& 0.38 (0.08 0.007)& 0.39 (0.09 0.004) \\
		\bottomrule
	\end{tabular}
	% Or to place a caption below a table
	\caption{The mean Hurst exponent $\overline{H}$ with (standard deviation, standard error) for all cell categories.}
	\label{tab:hurst}
\end{table}%   

\paragraph{Autocorrelation}
The anti-persistence can be further explored by considering the autocorrelation of the OCT4 time series. The autocorrelation is the correlation of a time series with itself at increasing time lags, hence $-1\leq C\leq1$ where $C=0$ signifies no correlation, $C<0$ a negative correlation and $C>0$ a positive correlation. The decay of the autocorrelation to zero (scaled to cell lifetimes) is presented in the Appendix in Ref. \cite{Wolff18} and here we extend this to quantify the periods of anti-correlation and consider the periodic nature of the autocorrelation. 

Typical autocorrelation functions are shown in Figure~\ref{fig:regulation}(b). The majority of the cells follow an autocorrelation similar to the one shown in Figure~\ref{fig:regulation}(b)i (Cell ID 46), with initial correlation declining to zero, followed by a period of anti-correlation before the autocorrelation settles at zero. There are, however, other behaviours evident. Some cells show several lag intervals of anti-correlation, as in Figure~\ref{fig:regulation}(b)ii (Cell ID 14), with others showing a positive correlation at a longer time lag before settling at zero, as in Figure~\ref{fig:regulation}(b)iii (Cell ID 43).

Anti-correlation for a time lag of at least one hour duration is seen in 99\% (1255/1274) of cells, and for at least five hours in 86\% (1090/1274) of cells. Of the cells with at least one hour anti-correlation visible, 44\% show a second period of correlation near the end of their lifetimes (as in Figure~\ref{fig:regulation}(b)ii and iii). Out of these, $65\%$ show one period of anti-correlation (as in Figure~\ref{fig:regulation}(b)iii), $31\%$ two periods (as in Figure~\ref{fig:regulation}(b)ii), and the remaining 4\% three or more. For the 57\% of cells with no second period of correlation, 90\% of cells show one period, 8\% two periods and 2\% three or more periods of anti-correlation. There is no correlation between the number of periods of anti-correlation and cell fate or the cell's average position in the colony.  

The earliest time anti-correlation occurs, $t_{\rm{AC}}$, can be calculated for each individual cell. The distribution of $t_{\rm{AC}}$ for cells with at least one hour anti-correlation is shown in the Supplementary Information, Figure S7, showing that in all cells anti-correlation has begun by 8 hours into the cell cycle. This could be due to the memory effects or the down-regulation of the PTF which occurs prior to mitosis \cite{Zaret14,Festuccia17}. The distributions of the percentage lifetime a cell spends in an anti-correlated state are given in Figure S7. Across all groups cells spend 40-80\% (with a mean of 60\%) of their cell cycle expressing anti-correlated OCT4.

The oscillatory nature and decay of the autocorrelation can be captured by the function $C=\cos(2\pi t/a)e^{-t/b}$ \cite{Sveshnikov66}, where $a$ and $b$ are constants and $t$ is time (note that this periodicity in the autocorrelation does not necessarily imply periodicity in the time series). These fittings are shown in Figure S8 for 25 randomly selected cells. This quantifies the temporal, periodic decay in the autocorrelation, with the parameter $a$ representing the time-scale of the periodicity, and $b$ the time-scale of the decay (the correlation decay time). Histograms of $a$ and $b$ for all 1274 cells are shown in Figure S9 and S10. For all fates, the medians are 11.7\,h and 3.0\,h, with 90th percentiles of 30\,h and 7\,h for $a$ and $b$ respectively. This quantifies the characteristic time-scale of the periodicity and the correlation decay time as less than 7 hours in 90\% of cases.
  
The correlation time is defined as ${\tau}=\int_{-\infty}^{\infty}C(t) dt$, with a mean across all cells of $\overline{\tau}\approx 0 \pm 0.002\,$h. The distribution of all $\tau$ is shown in Figure S7. The mean autocorrelation (Figure S11) decreases to zero at around three hours, followed by a period of negative autocorrelations between approximately three and 12 hours. By 13 hours, the mean autocorrelation settles at zero, showing no internal memory past this time. These observations are robust to cell fate and the equivalent autocorrelations for pluripotent and differentiated cells are shown in Figure S12. This shows that during a cell cycle, there is long-term memory in the OCT4 expression up to around 12 hours, but the nature of the effect differs over this time with initial correlation being replaced by anti-correlation. Notably, the mean autocorrelation is not fully described by $\cos(2\pi t/a)e^{-t/b}$, as the full scale of the anti-correlation is not captured.

\paragraph{Diffusion analysis}
The theory of diffusion and random walks is widely used across many biological applications, including stem cells and so it is useful to quantify the OCT4 behaviour within this framework \cite{Murray02,Codling08,Li10,Wu14,Wadkin17,Wadkin18}. After the cell division with asymmetric inheritance of OCT4 there is a short period of increased fluctuations \cite{Wolff18}. Therefore here we consider each OCT4 time series from half an hour after cell division. The mean square difference of OCT4 over time, MSD$(t)$, can be calculated as $\langle |\Omega(t_i)-\Omega(t_0)|^2 \rangle$, where the angular brackets denote the average across all cells in the group considered. 

\begin{figure*}[h]
	\centering
	\begin{subfigure}[t]{1\textwidth}
		\includegraphics[width=\textwidth]{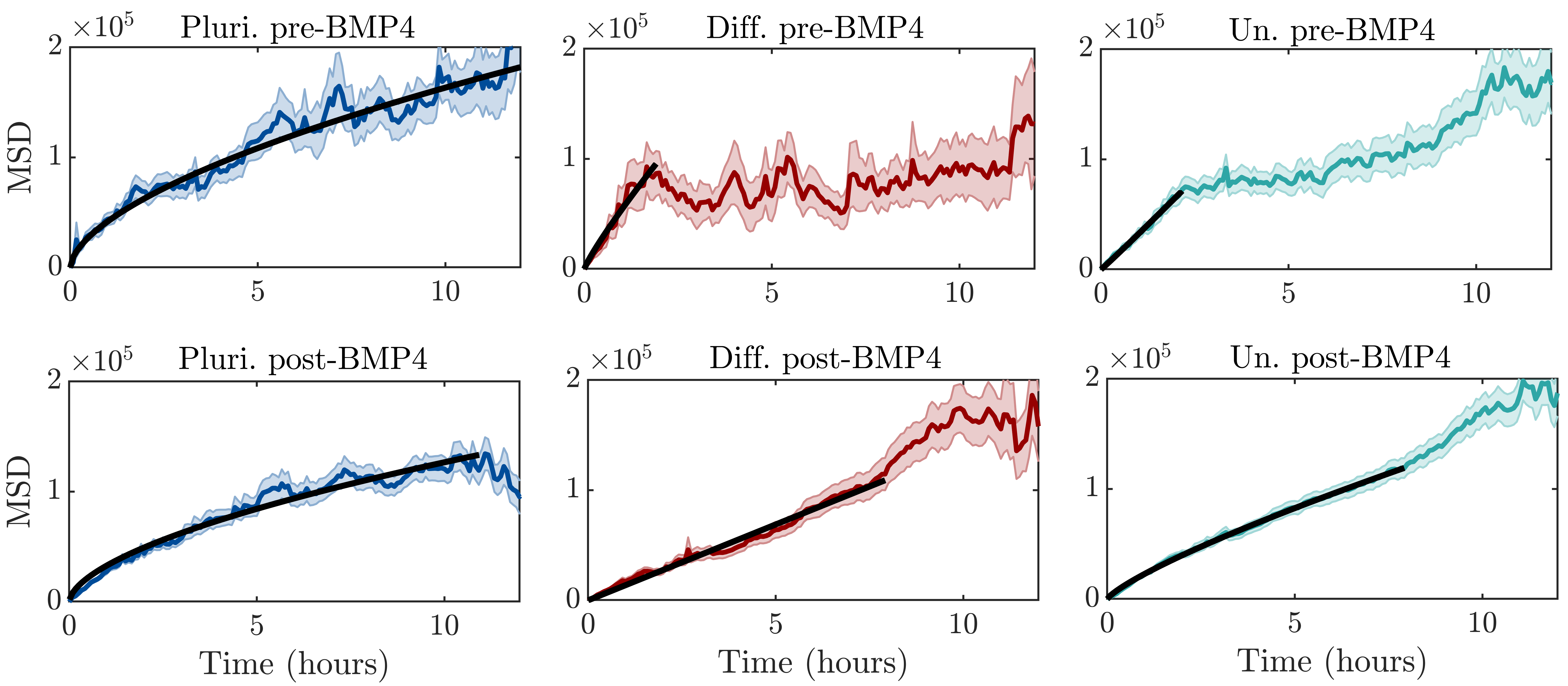}
	\end{subfigure}	
	\caption{\label{fig:msd}The mean-square difference up to 12 hours for each pro-fate, pre- and post-BMP4. The black lines show the power law fits, MSD$=\beta t^\alpha$, with the parameters $\alpha$ and $\beta$ specified in Table~\ref{tab:diff} and \ref{tab:diff2}.}
\end{figure*}

The MSD for each pro-fate, pre- and post-BMP4 between 0 and 12 hours is shown in Figure~\ref{fig:msd}. In all cases, there is visible sub-diffusive behaviour. The power law fits $\mathrm{MSD} =\beta t^\alpha$ are shown in Figure~\ref{fig:msd} and the parameters $\alpha$ and $\beta$ are given in Table~\ref{tab:diff} and \ref{tab:diff2}. This sub-diffusivity is consistent with the anti-persistence. The effect of this on sister cell's OCT4 is presented in the Supplementary Information, Figure S13 and S14. The differentiated and unknown (pre-BMP4) cells have $\alpha\approx1$ showing diffusion at early times. %***note this measure how far away is the current oct4 level from its start point. are we getting further away from the initial value faster or slower than you would expect if the series was behaving randomly. So for post-bmp4, the differentiated cells show a strong reduction in expression (i.e. an increase in the their distance from the initial point) and therefore the MSD is less suppressed. The lack of evidence of superdiffusivity interesting? **

\begin{table}[h]
	\centering
		\begin{tabular}[t]{ccc}
			\toprule
			$\alpha$ & Pre-BMP4 & Post-BMP4\\
			\midrule
			P & $0.59\pm0.03$  & $0.58\pm0.03$ \\
			D & $0.88\pm0.13$  & $1.00\pm0.004$ \\
			U & $1.01\pm0.05$ & $0.80\pm0.01$ \\
			\bottomrule
		\end{tabular}

	% Or to place a caption below a table
	\caption{The parameter $\alpha$ in the fitting to the mean square displacement with time, $\mathrm{MSD} =\beta t^\alpha$ for all cell pro-fates.}
	\label{tab:diff}
\end{table}%   

\begin{table}[h]
	\centering
	\begin{tabular}[t]{ccc}
		\toprule
		$\beta$ & Pre-BMP4 & Post-BMP4\\
		\midrule
		P & $42000\pm2700$ & $33000\pm2000$ \\
		D & $54000\pm3000$ & $13000\pm900$ \\
		U & $32000\pm1000$ & $23000\pm400$ \\
		\bottomrule
	\end{tabular}
	\caption{The parameter $\beta$ in the fitting to the mean square displacement with time, $\mathrm{MSD} =\beta t^\alpha$ for all cell pro-fates.}
	\label{tab:diff2}
\end{table}%   

This further quantifies the self-regulatory behaviour of OCT4 within the diffusion framework, a fundamental starting point for many mathematical models. In the next sections we analyse the fate transitions and spatial patterning within the colony.

\subsection*{Fate transitions}

How a stem cell divides to give rise to two daughters is critical for the maintenance and expansion of the culture. Stem cells may undergo both symmetric and asymmetric cell divisions, guided by several molecular, cellular, and environmental cues \cite{Chacondev18,Saha08}. During symmetric cell division, a stem cell generates two stem cells or two differentiated cells. The former is highly desired as it leads to the maintenance of the pluripotent state in a hESCs colony. Asymmetric divisions result in only one daughter inheriting the fate of the mother cell \cite{Shahriyari13}. This is the main process driving the homeostatic growth of tissues in an organism \cite{Yamashita2010}. The quantification of the transition dynamics between the cell fates is of utmost importance to emulate this behaviour with experiments {\it in-silico}.

The classification of the cells in terms of their pro-fates at the end of the experiment allows us to study the transitions between the different pro-fates at mitosis \cite{Wolff18}. We define our notation as follows: let $m$ be the fate of the mother cell and $d_1$, $d_2$ the fates of its two daughters. We then indicate with $\left[m, d_1, d_2\right]$ the six possible outcomes for the two daughter cells, with $m$, $d_1$ and $d_2$ taking any of the three pro-fates, that is, P (pluripotent), U (unknown) and D (differentiated). Using the family trees, we calculate the transition probabilities between these pro-fates, shown in the Supplementary Information, Figure S15, Table S3 and S4. In the absence of BMP4, the most important event driving the fate dynamics in the colony are symmetric cell divisions with both daughters having the same state as their mother cell (denoted as [P, P, P], [D, D, D] and [U, U, U]).

A quantitative way of visualising these results is using one-step transition probability matrices that govern the changes between the pro-fates.  We use a state-transition Markov model as a simple tool to simulate the pro-fates (cohorts) long-term states in the colony. Note that we are not simulating the behaviour of single-cells with well-defined (correlated) transitions over their life-times. But instead focusing on a group of cells belonging to a cohort (that is P, U, D) for which the transitions between each other follow a Markov process \cite{Iskandar18}. Thus, the following calculations are not in contradiction the anti-persistent behaviour or the results presented by \cite{Wolff18} with the cell history being predictive of the cell fate. 

Since the presence of BMP4 significantly alters the underlying dynamics of the OCT4 signal, we hypothesise that a similar effect might influence the transition between the hESCs pro-fates. Thus we obtain two right stochastic matrices, the first showing the transitions for mother cells born before (absence) BMP4 treatment (303 events),
\begin{equation}
W =\bordermatrix{~ & \mathrm{P} & \mathrm{U} & \mathrm{D}\cr
	P & \textbf{0.603}  & 0.346 & 0.051 \cr
	U & 0.358  & \textbf{0.512} & 0.130\cr
	D & 0.294   & \textbf{0.368} & 0.338\cr},
\label{eq:matrixW}
\end{equation}
and the second showing the transitions for mother cells born after (presence) of BMP4 treatment (194 events),
\begin{equation}
W^* = \bordermatrix{~ &  \mathrm{P} & \mathrm{U} & \mathrm{D}\cr
	\mathrm{P} & \textbf{0.792}   & 0.172 & 0.036\cr
	\mathrm{U} & 0.367  & \textbf{0.513} & 0.120\cr
	\mathrm{D} & 0.000   & 0.304 & \textbf{0.696}\cr}.
\label{eq:matrixWstar}
\end{equation}
In Eq.~(\ref{eq:matrixW}) and (\ref{eq:matrixWstar}), the rows and columns correspond to transitions from the starting (mother cell) to ending (daughter cell) states, respectively. The events with the maximum probabilities are highlighted in bold. For $W$, Eq.~(\ref{eq:matrixW}), we get higher probabilities for $W_\mathrm{PP}$ and $W_\mathrm{UU}$. That is, if the mother is pluripotent, it has 60\% of probability of dividing into a pluripotent daughter, with a remaining  35\%  and 5\% of giving rise to an unknown and differentiated daughter, respectively. This last event, although small, is detrimental to the maintenance of a highly pluripotent colony. 

\begin{figure}
	\includegraphics[width=\columnwidth]{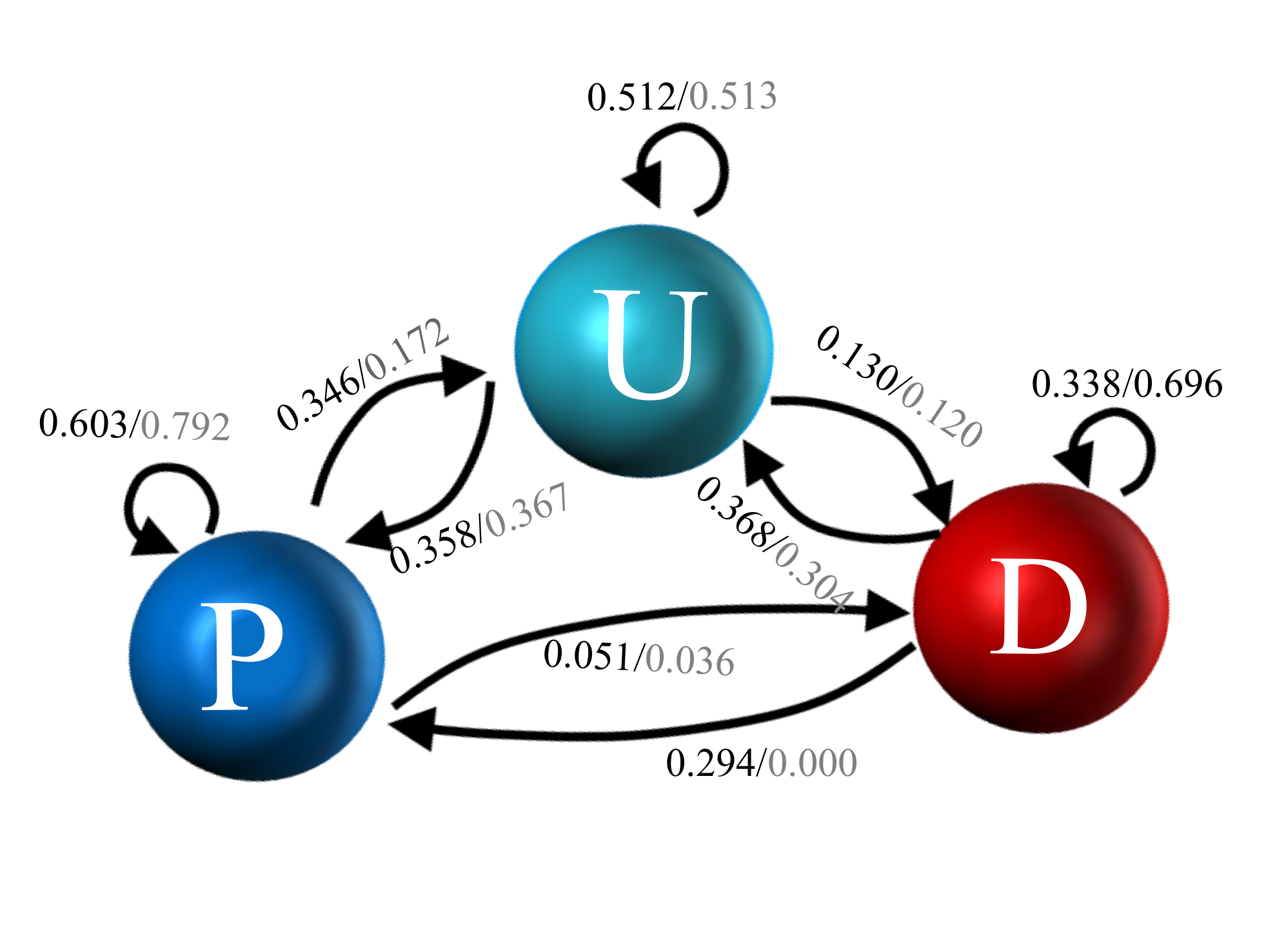}
	\caption{\label{fig:proFatesBefore} Transition probabilities between the pro-fates in hESCs in the absence/presence of BMP4. The pluripotent, differentiated and unknown pro-fates are represented with a blue, red, and turquoise spheres.}
\end{figure}

The one-step transition probabilities under BMP4 treatment show a strong tendency for all cell pro-fates ($W^*_\mathrm{PP}$, $W^*_\mathrm{UU}$ and $W^*_\mathrm{DD}$) to give rise to a daughter of the same pro-fate. The differentiation conditions changed from $W_\mathrm{DD} \sim 34\%$ to $W^*_\mathrm{DD} \sim 70\%$. That is, the pro-differentiated post-BMP4 mothers have twice the chance of producing a differentiated daughter cell than those pre-BMP4. Most importantly, our results reflect the fact that BMP4 treatment inhibits the return of a differentiated cell towards the pluripotent pro-fate, with $W^*_\mathrm{DP}\sim 0\%$. A scheme of these transition probabilities is shown in Figure~\ref{fig:proFatesBefore}. 

Next we consider the steady-state probability distributions, which quantify the averaged behaviour of the system towards a stationary state and how this state changes under a perturbation, such as the treatment of BMP4. Similar calculations have been applied to gene regulatory networks \cite{Li12}. We obtain the convergence of Eq.~(\ref{eq:matrixW}) and (\ref{eq:matrixWstar}) to a steady state using eigendecomposition. The following right-hand eigenvectors correspond to these stationary distributions,
\begin{equation}
\pi^{W} = \left[ 0.464, \, 0.418, \, 0.118 \right],
\label{eigen1}
\end{equation}
for transitions in the absence of BMP4 and
\begin{equation}
\pi^{W^\ast} = \left[ 0.523, \, 0.297, \, 0.180 \right],
\label{eigen2}
\end{equation}
in the presence of BMP4. These eigenvectors are also known as the stationary probability vectors of $W$ and $W^\ast$, respectively. After a sufficiently long time, the states dictated by Eq.~(\ref{eq:matrixW}) and (\ref{eq:matrixWstar}) will evolve towards a stationary probability distribution given by Eq.~(\ref{eigen1}) and (\ref{eigen2}). That is, given an initial distribution of cells across the pro-fate states, Eq.~(\ref{eigen1}) and (\ref{eigen2}) give the equilibrium distribution generated by the life trajectory of a cohort of identical cells by repeated multiplications of the vector of population counts by the transition probability matrix. It is important to note that these equations give the approximate behaviour of a strongly idealised system, thus we are not considering all the biological events affecting each cell in the colony. These type of models have proven successful in clinical decision making as a prognostic tool to guide decision making \cite{Iskandar18}.

Eq.~(\ref{eigen1}) indicates that in the stationary state, a hESC colony maintained under self-renewal conditions (absence of BMP4), 46\% of the cell state transitions result into highly pluripotent hESCs with the remaining 42\% and 12\% giving hESCs in the unknown and differentiated pro-fate. This last quantity indicates that in the long-term, over one-tenth of the cells will be in a differentiated state. 

The treatment with BMP4 changes the equilibrium state. The population fractions ending in the pluripotent and differentiated pro-fates increase to 52\% and 18\% respectively. This at the expense of a decrease in the fraction of cells in the unknown pro-fate. Results presented in Ref. \cite{Rosowski15} show that hESC colonies under BMP4 treatment show bands of differentiation with constant width independent of the colonies' radii. A straightforward calculation shows that the number of cells in the band surrounding a colony increases linearly with its radius, while the number of cells in the bulk increases following a power law. Thus, if a similar process is affecting the hESCs differentiation, this also leads only to a slight increase in transitions towards the differentiated pro-fate.

A potential factor affecting these fate transitions is the interaction between neighbouring cells. This phenomenon is achieved through a variety of signalling pathways and can impact cell state changes \cite{LiDev19,Chen14,Blauwkamp2012}. Next, we compute the spatio-temporal fate segregation in the colony, which serves to explain the high likelihood of certain transitions in Eq.~(\ref{eq:matrixW}) and (\ref{eq:matrixWstar}). 

\subsection*{Fate segregation}

The segregation of cells in the early mammalian embryo occurs during the early phases of embryonic development and ends with the formation of the three germ layers \cite{Kurosaka08}. The continuous rearrangement of cells occurs due to changes in the environment (surface cues) that induce differences in adhesion properties and changes in the cytoskeleton \cite{Fagotto3303}. These differences in adhesion properties between neighbouring cells maintain a physical separation between different cell types and are one of the basic mechanisms for the pattern formation during development \cite{KRENS2011189}.  

We use computational tools previously introduced in Ref. \cite{SirioQuantification} to quantify the segregation of the hESCs in terms of their pro-fates. We identify the set of nearest neighbours of each cell within the colony by applying the Voronoi tessellation diagram (VD) of the space to each snapshot of the colony. The state of the hESC colony at $t_\mathrm{exp} = 18\,$h is shown in the Supplementary Information,  Figure S16. Using the VD and its dual, the Delaunay triangulation, we can obtain a cell's set of nearest neighbours. The segregation of two types of cells, A and B, can be measured using the segregation order parameter, $\delta$, that depends explicitly on the number of nearest neighbours \cite{SirioQuantification}. For a perfectly mixed system, with six Delaunay neighbours, $\delta \sim 0$. If the system is completely segregated (e.g., one cluster of A particles surrounded by other of B particles) $\delta \sim 1$. 

\begin{figure}
	\begin{center}
		\includegraphics[width=\columnwidth]{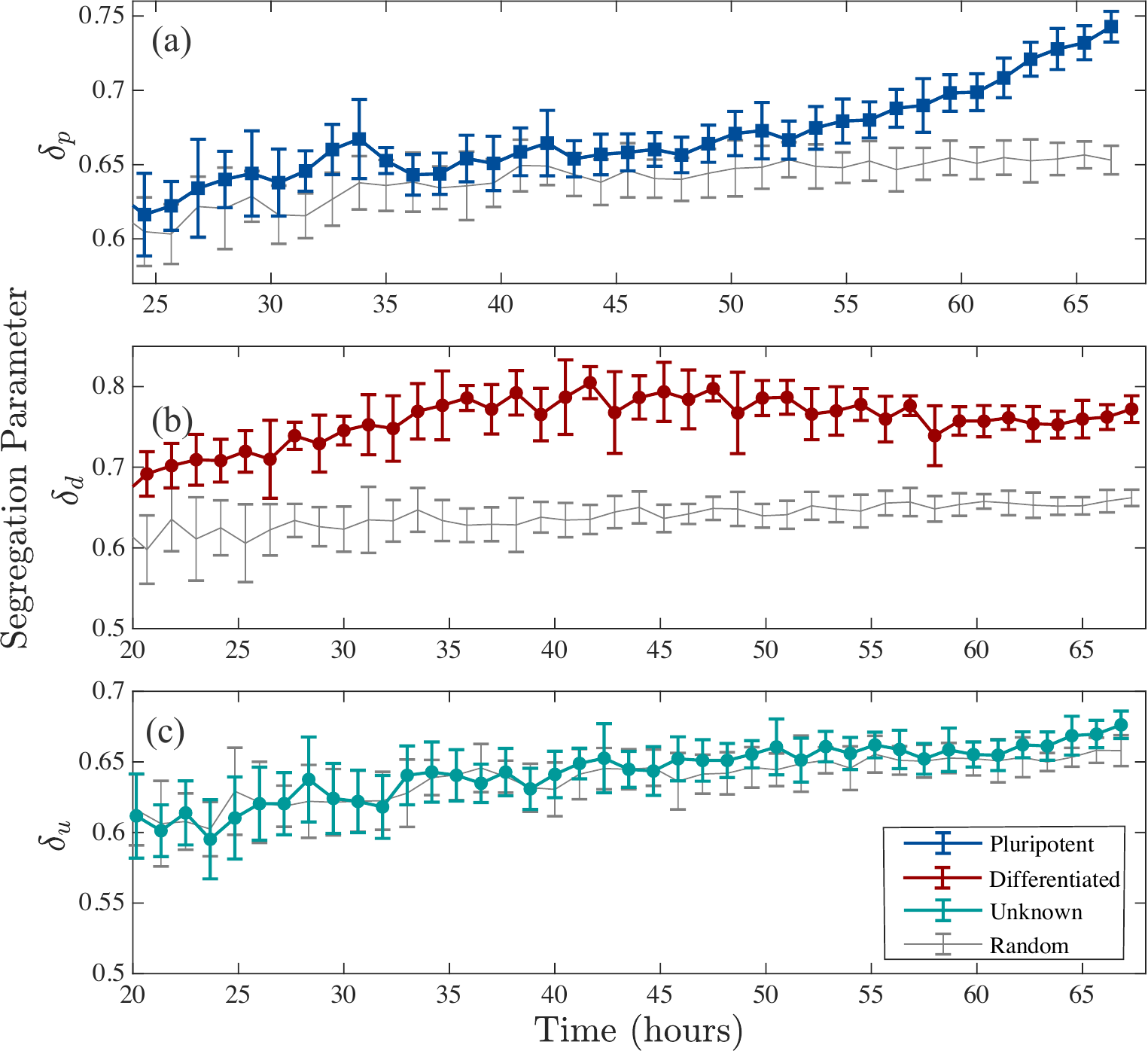}
	\end{center}
	\caption{Segregation of hESCs according to their pro-fates: (a) pluripotent ($\delta_\mathrm{P}$) (b) differentiated ($\delta_\mathrm{D}$)  and (c) unknown ($\delta_\mathrm{U}$) cells as a function of time. The results of the bootstrap method for each sample are shown in grey which correspond to the calculations performed by re-sampling the datasets with replacement. Each data point shows the average value with standard deviation error bars obtained by averaging over all cells in 12 snapshots (one hour of experiment).}
	\label{fig:SegregationParameter}  
\end{figure}

For this calculation, the segregation of the pluripotent (or differentiated) cells is obtained by merging the unknown cells with the differentiated (or pluripotent) cells to generate the type B cells. The results are shown in Figure~\ref{fig:SegregationParameter}. We discard the data at the initial stages of the experiment to avoid spurious results due to the low cell numbers and poor statistics. The hESCs with pluripotent pro-fate are effectively segregated ($\delta_\mathrm{P} > 0.65$) from the other pro-fates after $t_\mathrm{exp} \approx 48\,$h  (two days after the beginning of the experiment), as seen in Figure \ref{fig:SegregationParameter}(a) and Video S1. The differentiated pro-fate, Figure~\ref{fig:SegregationParameter}(b), segregates ($\delta_\mathrm{D} > 0.7$) at earlier stages, $t_\mathrm{exp} \approx 20\,$h (one day after the beginning of the experiment), and remain in that state.  The result for the unknown pro-fate (with type B cells defined as both the differentiated and pluripotent hESCs) is inconclusive. This is expected since these cells are located between the pluripotent and differentiated pro-fates and thus, are `mixed' with their type B cells. 

\subsection*{OCT4 dissimilarity}

The measurement of the clustering according to the hESCs pro-fates using the segregation order parameter allows us to calculate the OCT4 variability between a specific hESC and its closest neighbours. We use the set of neighbours obtained for each hESCs in the previous section and characterise the OCT4 levels between the cells located in a local neighbourhood. Since the OCT4 levels take any real positive value, we define the `cooperation' between the cells as the tendency of a specific cell to express a similar OCT4 value to that of its nearest neighbours.

We use the dissimilarity metric, $S_i$, to quantify the similarity in OCT4 in neighbouring cells \cite{Okamoto}. If the OCT4 values in the colony are quantitatively more similar amongst adjacent cells, $S_i \approx 0$. If the opposite occurs, that is, the OCT4 values are different between the cells in a local neighbourhood, $S_i \neq 0$. 

\begin{figure*}[!]
	\begin{center}
		\includegraphics[width=1\textwidth]{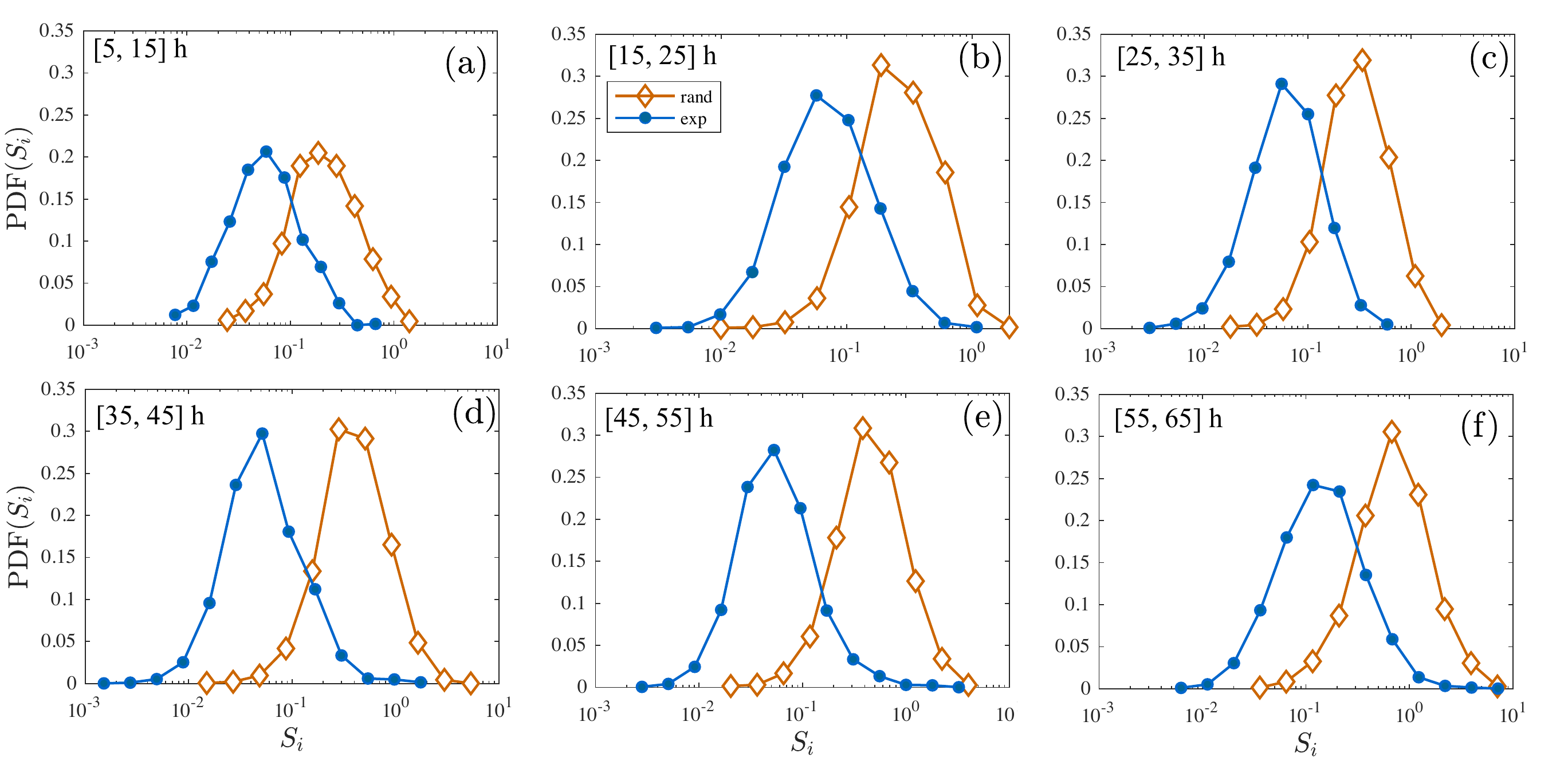}
	\end{center}
	\caption{Probability density functions for the dissimilarity metric $\langle S_i \rangle$ of the OCT4 expression of hESCs and their nearest neighbours (filled blue circles). Simulated datasets with the OCT4 values following a uniform distribution are shown with open orange diamonds.}
	\label{fig:DissimilarityMetric}  
\end{figure*}

To avoid inaccurate results due to poor statistics, we obtain the probability density function (PDF) of $S_i$  over specific time intervals, Figure~ \ref{fig:DissimilarityMetric}.  We also perform a qualitative comparison with simulated datasets, by randomising the positions of the cells and drawing their OCT4 levels from a uniform distribution over the same range as the experimental distribution, assuming that no cell-to-cell interactions occur. To assist in the visualisation of these plots,  we use a non-linear binning scheme and plot the $x$-axis on a logarithmic scale.

These results indicate a similar behaviour in the PDFs of $S_i$ in the absence of BMP4, that is, for $t_\mathrm{exp} = \left[5, 35 \right]\,$h, Figure~\ref{fig:DissimilarityMetric}(a-c), with an average mean of $0.081 \pm 0.009$ and skewness of $3.07 \pm 0.56$. For $t_\mathrm{exp} = \left[35,45\right]\,$h and $\left[45,55\right]\,$h, panels (d) and (e), the mean of these two distributions is $0.078 \pm 0.005$, similar to those values observed for (a-c). However, their tails become larger, with a skewness of 6.90 and 9.84, respectively. These latter results indicate that some (few) cells in the colony are expressing highly dissimilar OCT4 values compared with those of their nearest neighbours. These differences become larger for $t_\mathrm{exp} = \left[45,55\right]\,$h, after the treatment with BMP4. Finally, for  $t_\mathrm{exp} = \left[56,65 \right]\,$h, Figure \ref{fig:DissimilarityMetric}(f), the distribution displaces towards the right (with a mean of $0.2268$ and skewness of 6.74). This corresponds to an overall increment in the dissimilarity for a large proportion of the cells in the colony. In all cases, the PDFs obtained with the simulated data have a larger mean and smaller skewness than the experimental datasets. 

\section{Discussion}

We set out to demonstrate transferable approaches for quantifying the dynamics of pluripotency transcription factors in hESC colonies, using the published OCT4 data set from Ref. \cite{Wolff18}.

\subsection*{Temporal regulation}

%Time-lapse experiments such as Ref. \cite{Wolff18} allow the quantification of temporal TF regulation which can be compared to, and enhance, current biological knowledge. For example, a sharp decline in OCT4 levels occurring before cell division is noted in Ref. \cite{Wolff18}, in keeping with the transcription factor down regulation known to occur before mitosis \cite{Zaret14,Festuccia17}. This phenomena can be quantified, with the decrease in OCT4 beginning, on average, 35 minutes before cell division, lasting for 15 minutes, and showing a reduction of 22\% from the interphase OCT4 expression. This is shown in the Supplementary Information, Figure S14. The OCT4 expression levels recover as mitosis occurs and the cycle repeats for the cell's descendants consistent with experimental results showing OCT4 resets on re-entry to the G1 phase \cite{Shin16,Kim18}. 

Sister cells show more closely related OCT4 values than pairs of random cells \cite{Wolff18}. Here we have further quantified the temporal dynamics of sister cells in relation to one another. Taking into account any common trends affecting both cells due to their shared environment, the sister cells pre-BMP4 show a moderate correlation with each other (0.5). This is reduced to a slight correlation for pairs that both exist post-BMP4 (0.3). The fact that these correlations still occur after de-trending further highlights the inherent similarities between sister cells. 

We found that pre-BMP4, within the first 24 hours of the colony growth, all hESCs with pluripotent pro-fate express high levels of OCT4. The averaged expression of these signals evolves in time towards a minimum coinciding with the treatment with BMP4, and with strong correlations between the pluripotent and unknown pro-fates (0.84). This correlation may indicate the presence of an underlying process affecting the behaviour of the cells at the colony level, that is, changes in the environmental conditions, cell-to-cell signalling, paracrine signals, etc., that may have influenced the colony at the early stages. The OCT4 distribution at the colony level is not uniform, in contrast with the results observed for other ESCs types \cite{Navarro12,Munoz12}, showing a highly dynamical behaviour and evolving from higher to lower values of OCT4 expression as the colony grows.

The time-step changes in OCT4 are symmetric, with an average of zero, best described by a Laplace distribution. Laplace distributions have previously been applied to gene expression data, with the suggestion that the distribution can represent mixtures of other distributions (e.g., Normals, Pareto) also related to gene expression \cite{Purdom05}. The parameters can be used for experimental comparisons, as direct inputs into computational models and for model verification. Further experimental data (e.g., different cell lines, culture conditions, restricted geometries, cell densities) are needed confirm the robustness, estimate the parameters for other experimental conditions and investigate how this is affected by cell-cell interactions. Such stochastic fluctuations in OCT4 have been shown to bias cell fate \cite{Strebinger19} with evidence of asymmetric noise leading to noise-mediated cell plasticity \cite{Holmes17}. 

Although this shows that overall, positive changes in OCT4 are just as likely to occur as negative ones, it does not reveal anything about the temporal nature of these fluctuations and hence any temporal correlation properties (for example, all the positive changes in OCT4 could come one after the other, followed by all the negative changes, it does not mean that a positive change is necessarily followed by a negative change). There is also a difference in these fluctuations after the differentiation agent, with the addition of BMP4 provoking tighter self-regulation across all cell fates. Further experiments with increasing colony size are needed to investigate whether this self-regulation is a collective behaviour effect. 

A significant finding is the quantification of the self-regulatory properties of OCT4. We calculate the average Hurst exponent (0.38, indicative of anti-persistence) which has been previously used to characterise gene expression \cite{Ghorbani18}, DNA sequencing \cite{Liu15}, stem cell division times \cite{Bogdan14} and self-renewal capacities \cite{Sieburg13}. Broadly, the identification of a Hurst exponent which is not Brownian suggests the use of specific equations for describing temporal OCT4. Furthermore, it can be a direct input into some stochastic modelling equations in which the parameter, $H$, is required, e.g., fractional Brownian motion \cite{Mandelbrot68,Mielniczuk07,Lacasa09,Barunik10}.
To visualise the OCT4 values in relation to the previous time frame we use Poincar\'e plots, but note that other methods are also applicable here, such as diffusion maps \cite{Haghverdi15}. 

An autocorrelation analysis shows periods of anti-correlation, in keeping with the regulation of pluripotency TFs \cite{Li04,Wang12,Akberdin18}. Throughout the colony growth, anti-correlation of at least five hours is seen in 86\% of cells (with no significant difference between the cell fates) and on average occurs between 3 and 12 hours into a cell's lifetime. This suggests that the anti-correlation is an inherent property of the cells, across all cell fates. Further experiments are needed to clarify that this is the case under different experimental conditions (i.e., different sized colonies, other cell lines, in different geometries) but this provides a further quantitative statistic for comparisons. The identification of this systematic property has implications for the underlying stochastic chemistry of the OCT4 regulation and can be used to inform chemical models of TF regulation, often based on the Hill equations \cite{Chickarmane06,Likhoshvai07}. 

The sub-diffusive nature of the time series allows for another characterisation of the OCT4 self-regulation using only two summary parameters, $\alpha$ and $\beta$. Note that this is not contradictory to the conclusion of pre-determined cell fate in Ref. \cite{Wolff18}, as here we consider the behaviour over individual cell lifetimes, (i.e., shorter time scales) and do not take into account other behaviours in the colony (i.e., over multiple cell divisions, longer time scales). The sub-diffusivity highlights the presence of a universal behaviour in the cells, and can be used as a direct parameter input into Brownian (and the fractional and geometric extensions) computational models allowing the future behaviour to be predicted statistically.

\subsection*{Fate transitions}

The one-step transition probability matrices depict the possible paths for cellular pro-fate transitions and give insight into the (a)symmetric nature of the mitotic events. Consistent with the results reported in Ref. \cite{Wolff18}, the pro-pluripotent hESCs have a higher probability ($\sim 60\%$) of giving rise to a daughter cell of the same fate. The remaining probability is associated with a pro-pluripotent cell giving rise to an unknown or differentiated daughter cell at mitosis. Interestingly, the pro-differentiated cells had the highest chance of giving rise to an unknown pro-fate cell. The transitions away from the pluripotent pro-fate are detrimental for the maintenance of a pluripotent colony if we assume that both pro-fates (unknown and differentiated) are undesirable for highly pluripotent colonies. 

Under the framework of Markov chains, we obtained the stationary probability functions of hESCs fate transitions both in the absence and presence of BMP4. The stationary probability distributions predict 46\% of the cell state transitions towards a cell in the pluripotent pro-fate for a hESC colony under self-renewal conditions and in the absence of BMP4. The remaining 42\% and 12\% result in hESCs in the unknown and differentiated pro-fates, respectively. Thus, after a sufficiently long time, the proportion of hESCs moving to the pluripotent and unknown pro-fates becomes similar. In the presence of BMP4, we obtain a stationary probability distribution with 52\%, 30\% and 18\% of cell state transitions towards the pluripotent, unknown and differentiated pro-fates. As expected, this last probability indicates that BMP4 induces an increase in the number of hESCs transitioning to a differentiated state. 

%We hypothesise that this is a result of several factors acting in conjunction, for example, pluripotent hESCs have a short cell-cycle \cite{Calder13,Neganova_NANOG,Neganova2009} which lengthens when they differentiate. This directly affects the rate of occurrence of cell transitions towards the differentiated pro-fate. Furthermore, following the results presented in Ref. \cite{Rosowski15} for hESC colonies under BMP4 treatment, this decline in transitions towards the differentiated state may occur to accommodate the bands of differentiation with constant width ($149 \pm 59 \mu$m) observed in colonies of different radii.

\subsection*{Spatial segregation}

The segregation parameter indicates that the ancestors of the hESCs with a differentiated pro-fate position themselves at the outer (top) regions of the colony as early as one day after the beginning of the experiment and they remain clustered (segregated) throughout the experiment. The segregation process culminates with the separation of the pluripotent and unknown pro-fates one day later. Coupling these results with the higher probabilities of division towards a daughter with the same pro-fate, a hESCs within a differentiated pro-fate gives rise to a differentiated daughter in its local neighbourhood. These transitions are consistent with the patterning observed in hESCs under confinement inside microfluidic devices reported elsewhere \cite{Warmflash14,Tewary4298}. 

The dissimilarity metric indicates that the OCT4 values amongst the nearest cells remain comparable in the absence of BMP4. However, we observe changes in the tail of these distributions 10 hours after treatment with BMP4. This indicates the presence of cells expressing highly dissimilar OCT4 values with those of their nearest neighbours. Since by this time the pluripotent and differentiated pro-fates are segregated, we hypothesise that these large differences in OCT4 levels may happen at the interface between the differentiated and unknown pro-fates since their OCT4 distributions are highly dissimilar by the end of the experiment.

\subsection*{The unknown fate}

The behaviour of the unknown cells (unable to be classified as either pluripotent or differentiated based on their OCT4 and CDX2 levels) lies between that of the pluripotent and differentiated cell fates. Spatially, they are not clearly segregated from the pluripotent and differentiated cells, a possible indication that the fate decision has not yet occurred. The unknown pro-fate cells could be the result of a mixture of both populations with the ability to express high and low OCT4 expressions, or cells undergoing a transition phase between pluripotent and differentiated. The distinct differences between cell fates could provide non-invasive diagnostic tools to identify cell fates. 

\subsection*{Future}

The experiment in Ref. \cite{Wolff18} has led to a rich analysis, allowing us to establish the language through which to quantitatively compare this experiment to others and to guide mathematical modelling choices. These tools can be easily modified and adapted to study the dynamics of other relevant transcription factors, such as NANOG and SOX2 in H9 hESCs. Furthermore, for experiments in which several transcription factors are measured simultaneously in real-time, our custom-developed algorithms can be easily modified to quantify the \textit{in-vivo} segregation in tissues. 

The limitation of this research is the single hESC colony analysed. Although several mitotic events were recorded during the experiment, the results need to be generalised to other hESC colonies of different cell lines and under different experimental conditions. Currently, methods such as single-cell RNA sequencing and real-time reverse transcription PCR allow more accurate quantification of the cellular states at the transcriptional level and gene expression, respectively. These datasets can potentially drive the development of mathematical models at the single-cell level using stochastic processes (e.g., the non-Markovian Langevin or non-Markovian Fokker-Planck equations). These models can account for transitions between cell states (genotype) and/or fates (phenotype) at two different time scales: shorter and longer than the cell cycle. A combined measurement of TF expression in hESCs with single-cell RNA sequencing may give a deeper understanding of the spatio-temporal changes of TF expression in hESC colonies.

In general, our results highlight the need for further temporal experimental data on OCT4 and other transcription factors. We expect that studies with computational tools complementing experiments will become more commonplace, furthering our knowledge in stem cell biology and accelerating the development of stem cell-based technologies.

\section*{Author Contributions}

Methodology: L.E.W, S.O-F., R.A.B., A.S., N.G.P.; Software: S.O-F.; Data analyses: L.E.W., S.O-F.;  Writing: L.E.W., S.O-F; Editing: L.E.W., S.O-F., I.N., M.L., R.A.B., A.W.B., N.G.P.; Supervision: R.A.B., A.S., N.G.P.

\section*{Acknowledgements}
SOF acknowledges financial support from the Consejo Nacional de Ciencia y Tecnología  (CONACyT, Mexico) for the grant CVU-174695. IN acknowledges the grant from the Russian Government 641 Program for the recruitment of the leading scientists into 641 Russian Institution of Higher Education 14.w03.31.0029 and RFFI project grant number 20-015-00060. ML acknowledges BBSRC UK (BB/I020209/1) for providing financial support for this work. RAB wants to acknowledge financial support from CONACyT (Mexico) through the project 283279.

\section*{References}
\bibliographystyle{unsrt}
\bibliography{refs}

\begin{thebibliography}{10}

\bibitem{TAKAHASHI2007861}
Kazutoshi Takahashi, Koji Tanabe, Mari Ohnuki, Megumi Narita, Tomoko Ichisaka,
  Kiichiro Tomoda, and Shinya Yamanaka.
\newblock Induction of pluripotent stem cells from adult human fibroblasts by
  defined factors.
\newblock {\em Cell}, 131(5):861 -- 872, 2007.

\bibitem{Li04}
M.~Li and J.~C. Izpisua~Belmonte.
\newblock {{D}econstructing the pluripotency gene regulatory network}.
\newblock {\em Nat. Cell Biol.}, 20(4):382--392, 2018.

\bibitem{DANIELS20103563}
Brian~R. Daniels, Christopher~M. Hale, Shyam~B. Khatau, Sravanti Kusuma,
  Terrence~M. Dobrowsky, Sharon Gerecht, and Denis Wirtz.
\newblock Differences in the microrheology of human embryonic stem cells and
  human induced pluripotent stem cells.
\newblock {\em Biophys. J.}, 99(11):3563 -- 3570, 2010.

\bibitem{Bauwens08}
C.~L. Bauwens, R.~Peerani, S.~Niebruegge, K.~A. Woodhouse, E.~Kumacheva,
  M.~Husain, and P.~W. Zandstra.
\newblock Control of human embryonic stem cell colony and aggregate size
  heterogeneity influences differentiation trajectories.
\newblock {\em Stem Cells}, 26(9):2300--2310, 2008.

\bibitem{Ebert10}
A.~D. Ebert and C.~N. Svendsen.
\newblock {Human stem cells and drug screening: opportunities and challenges}.
\newblock {\em Nat. Rev. Drug Discov.}, 9(5):367--372, 2010.

\bibitem{Zhu13}
Z.~Zhu and D.~Huangfu.
\newblock {Human pluripotent stem cells: an emerging model in developmental
  biology}.
\newblock {\em Development}, 140(4):705--717, 2013.

\bibitem{Avior16}
Y.~Avior, I.~Sagi, and N.~Benvenisty.
\newblock Pluripotent stem cells in disease modelling and drug discovery.
\newblock {\em Nat. Rev. Mol. Cell Biol}, 17(3):170–182, 2016.

\bibitem{Ilic17}
D.~Ilic and C.~Ogilvie.
\newblock {C}oncise {R}eview: {H}uman embryonic stem cells - what have we done?
  what are we doing? where are we going?
\newblock {\em Stem Cells}, 35(1):17--25, 2017.

\bibitem{Shroff17}
G.~Shroff, J.~D. Titus, and R.~Shroff.
\newblock A review of the emerging potential therapy for neurological
  disorders: human embryonic stem cell therapy.
\newblock {\em Am. J. Stem Cells}, 6(1):1, 2017.

\bibitem{Trounson16}
A.~Trounson and N.~D. DeWitt.
\newblock Pluripotent stem cells progressing to the clinic.
\newblock {\em Nat. Rev. Mol. Cell Biol}, 17(3):194, 2016.

\bibitem{Young05}
L.~A. Boyer, T.~I. Lee, M.~F. Cole, S.~E. Johnstone, S.~S. Levine, J.~P.
  Zucker, M.~G. Guenther, R.~M. Kumar, H.~L. Murray, R.~G. Jenner, D.~K.
  Gifford, D.~A. Melton, R.~Jaenisch, and R.~A. Young.
\newblock Core transcriptional regulatory circuitry in human embryonic stem
  cells.
\newblock {\em Cell}, 122(6):947—956, September 2005.

\bibitem{Chambers09}
I.~Chambers and S.~R. Tomlinson.
\newblock The transcriptional foundation of pluripotency.
\newblock {\em Development}, 136(14):2311--2322, 2009.

\bibitem{Pauklin13}
S.~Pauklin and L.~Vallier.
\newblock The cell-cycle state of stem cells determines cell fate propensity.
\newblock {\em Cell}, 155(1):135 -- 147, 2013.

\bibitem{Shuzui19}
E.~Shuzui, M.~Kim, and M.~Kino-oka.
\newblock Anomalous cell migration triggers a switch to deviation from the
  undifferentiated state in colonies of human induced pluripotent stems on
  feeder layers.
\newblock {\em J. Biosci. Bioeng.}, 127(2):246 -- 255, 2019.

\bibitem{Stadhouders19}
R.~Stadhouders, G.~J. Filion, and T.~Graf.
\newblock {{T}ranscription factors and 3D genome conformation in cell-fate
  decisions}.
\newblock {\em Nature}, 569(7756):345--354, 2019.

\bibitem{Nemashkalo17}
A.~Nemashkalo, A.~Ruzo, I.~Heemskerk, and A.~Warmflash.
\newblock Morphogen and community effects determine cell fates in response to
  bmp4 signaling in human embryonic stem cells.
\newblock {\em Development}, 144(17):3042--3053, 2017.

\bibitem{Rosowski15}
K.~A. Rosowski, A.~F. Mertz, S.~Norcross, E.~R. Dufresne, and V.~Horsley.
\newblock {{E}dges of human embryonic stem cell colonies display distinct
  mechanical properties and differentiation potential}.
\newblock {\em Sci. Rep.}, 5:14218, Sep 2015.

\bibitem{Hwang08}
N.~S. Hwang, S.~Varghese, and J.~Elisseeff.
\newblock Controlled differentiation of stem cells.
\newblock {\em Adv. Drug Deliv. Rev.}, 60(2):199 -- 214, 2008.
\newblock Emerging Trends in Cell-Based Therapies.

\bibitem{Warmflash14}
A.~Warmflash, B.~Sorre, F.~Etoc, E.~D. Siggia, and A.~H. Brivanlou.
\newblock {{A} method to recapitulate early embryonic spatial patterning in
  human embryonic stem cells}.
\newblock {\em Nat. Methods}, 11(8):847--854, 2014.

\bibitem{ZhanYuOCT4}
Zhen-Ning Zhang, Sun-Ku Chung, Zheng Xu, and Yang Xu.
\newblock {OCT4} maintains the pluripotency of human embryonic stem cells by
  inactivating p53 through sirt1-mediated deacetylation.
\newblock {\em Stem Cells}, 32(1):157 -- 165, 2014.

\bibitem{RodriguezRegulationOCT}
Ryan~T. Rodriguez, J.~Matthew Velkey, Carolyn Lutzko, Rina Seerke, Donald~B.
  Kohn, K.~Sue O’Shea, and Meri~T. Firpo.
\newblock Manipulation of {OCT4} levels in human embryonic stem cells results
  in induction of differential cell types.
\newblock {\em Experimental Biology and Medicine}, 232(10):1368 -- 1380, 2007.

\bibitem{Guilai2010}
Guilai Shi and Jin Ying.
\newblock Role of {OCT4} in maintaining and regaining stem cell pluripotency.
\newblock {\em Stem Cell Res. Ther.}, 1(5):39, 2010.

\bibitem{GuangjinThomson}
Guangjin Pan and James~A Thomson.
\newblock Nanog and transcriptional networks in embryonic stem cell
  pluripotency.
\newblock {\em Cell Research}, 17:42--49, 2007.

\bibitem{Lin18}
Yen~Ting Lin, Peter~G. Hufton, Esther~J. Lee, and Davit~A. Potoyan.
\newblock A stochastic and dynamical view of pluripotency in mouse embryonic
  stem cells.
\newblock {\em PLOS Computational Biology}, 14(2):1--24, 02 2018.

\bibitem{Wu12}
Jincheng Wu and Emmanuel~S. Tzanakakis.
\newblock Contribution of stochastic partitioning at human embryonic stem cell
  division to {NANOG} heterogeneity.
\newblock {\em PLoS One}, 7(11):1--14, 11 2012.

\bibitem{LAURENT2011106}
Louise~C. Laurent, Igor Ulitsky, Ileana Slavin, Ha~Tran, Andrew Schork, Robert
  Morey, Candace Lynch, Julie~V. Harness, Sunray Lee, Maria~J. Barrero, Sherman
  Ku, Marina Martynova, Ruslan Semechkin, Vasiliy Galat, Joel Gottesfeld, Juan
  Carlos~Izpisua Belmonte, Chuck Murry, Hans~S. Keirstead, Hyun-Sook Park, Uli
  Schmidt, Andrew~L. Laslett, Franz-Josef Muller, Caroline~M. Nievergelt, Ron
  Shamir, and Jeanne~F. Loring.
\newblock Dynamic changes in the copy number of pluripotency and cell
  proliferation genes in {H}uman {ESC}s and i{PSC}s during reprogramming and
  time in culture.
\newblock {\em Cell Stem Cell}, 8(1):106 -- 118, 2011.

\bibitem{Chambers14}
M-E Torres-Padilla and I.~Chambers.
\newblock Transcription factor heterogeneity in pluripotent stem cells: a
  stochastic advantage.
\newblock {\em Development}, 141(11):2173--2181, 2014.

\bibitem{HuangDev}
Sui Huang.
\newblock Non-genetic heterogeneity of cells in development: more than just
  noise.
\newblock {\em Development}, 136(23):3853–--3862, 2009.

\bibitem{Stewart2006}
Morag~H. Stewart, Marc Boss{\'e}, Kristin Chadwick, Pablo Menendez, Sean~C.
  Bendall, and Mickie Bhatia.
\newblock Clonal isolation of h{ESC}s reveals heterogeneity within the
  pluripotent stem cell compartment.
\newblock {\em Nature Methods}, 3(10):807--815, 2006.

\bibitem{Dakhore2018}
Sushrut Dakhore, Bhavana Nayer, and Kouichi Hasegawa.
\newblock Human pluripotent stem cell culture: Current status, challenges, and
  advancement.
\newblock {\em Stem cells international}, 2018:7396905--7396905, Nov 2018.

\bibitem{Stockholm07}
D.~Stockholm, R.~Benchaouir, J.~Picot, P.~Rameau, T.~M.~A. Neildez, G.~Landini,
  C.~Laplace-Builhé, and A.~Paldi.
\newblock The origin of phenotypic heterogeneity in a clonal cell population in
  vitro.
\newblock {\em PLoS One}, 2(4):1--13, 04 2007.

\bibitem{Li12}
W.~Li, L-B. Cui, and M.~K. Ng.
\newblock On computation of the steady-state probability distribution of
  probabilistic {B}oolean networks with gene perturbation.
\newblock {\em J. Comput. Appl. Math.}, 236(16):4067--4081, 2012.

\bibitem{Navarro12}
P.~Navarro, N.~Festuccia, D.~Colby, A.~Gagliardi, N.~P. Mullin, W.~Zhang,
  V.~Karwacki-Neisius, R.~Osorno, D.~Kelly, M.~Robertson, and I.~Chambers.
\newblock {OCT4}/{SOX2}-independent {N}anog autorepression modulates
  heterogeneous {N}anog gene expression in mouse es cells.
\newblock {\em EMBO J.}, 31(24):4547--4562, 2012.

\bibitem{Munoz12}
S.~Muñoz~Descalzo, P.~Rue, J.~Garcia-Ojalvo, and A.~M. Arias.
\newblock Correlations between the levels of {OCT4} and {N}anog as a signature
  for {N}aïve {P}luripotency in {M}ouse {E}mbryonic {S}tem {C}ells.
\newblock {\em STEM CELLS}, 30(12):2683--2691, 2012.

\bibitem{Wang12}
Z.~Wang, E.~Oron, B.~Nelson, S.~Razis, and N.~Ivanova.
\newblock Distinct lineage specification roles for {N}{A}{N}{O}{G}, {O}{C}{T}4,
  and {S}{O}{X}2 in human embryonic stem cells.
\newblock {\em Cell Stem Cell}, 10(4):440 -- 454, 2012.

\bibitem{Radzisheuskaya2014}
Aliaksandra Radzisheuskaya and Jos{\'e} C.~R. Silva.
\newblock Do all roads lead to {OCT4}?: The emerging concepts of induced
  pluripotency.
\newblock {\em Trends. Cell Biol.}, 24(5):275--284, May 2014.

\bibitem{Ovchinnikov}
D.~A. Ovchinnikov, J.~P. Turner, D.~M. Titmarsh, N.~Y. Thakar, D.~Choon Sin,
  J.~J. Cooper-White, and E.~J. Wolvetang.
\newblock Generation of a human embryonic stem cell line stably expressing high
  levels of the fluorescent protein m{C}herry.
\newblock {\em World J. Stem Cells}, 26(4):71--79, 2012.

\bibitem{Wolff18}
S.~C. Wolff, K.~M. Kedziora, R.~Dumitru, C.~D. Dungee, T.~M. Zikry, A.~S.
  Beltran, R.~A. Haggerty, J.~Cheng, M.~A. Redick, and J.~E. Purvis.
\newblock Inheritance of oct4 predetermines fate choice in human embryonic stem
  cells.
\newblock {\em Mol. Syst. Biol.}, 14(9):e8140, 2018.

\bibitem{Setty}
M.~Setty, V.~Kiseliovas, J.~Levine, A.~Gayoso, L.~Mazutis, and D.~Pe’ere.
\newblock Characterization of cell fate probabilities in single-cell data with
  {P}alantir.
\newblock {\em Nature Biotechnology}, 37:451--460, 2019.

\bibitem{Tewary4298}
Mukul Tewary, Joel Ostblom, Laura Prochazka, Teresa Zulueta-Coarasa, Nika
  Shakiba, Rodrigo Fernandez-Gonzalez, and Peter~W. Zandstra.
\newblock A stepwise model of reaction-diffusion and positional information
  governs self-organized human peri-gastrulation-like patterning.
\newblock {\em Development}, 144(23):4298--4312, 2017.

\bibitem{Babaie07}
Yasmin Babaie, Ralf Herwig, Boris Greber, Thore~C Brink, Wasco Wruck, Detlef
  Groth, Hans Lehrach, Tom Burdon, and James Adjaye.
\newblock Analysis of {O}ct4-dependent transcriptional networks regulating
  self-renewal and pluripotency in human embryonic stem cells.
\newblock {\em Stem cells}, 25(2):500--510, 2007.

\bibitem{Ghule11}
P.~N. Ghule, R.~Medina, C.~J. Lengner, M.~Mandeville, M.~Qiao, Z.~Dominski,
  J.~B. Lian, J.~L. Stein, A.~J. van Wijnen, and G.~S. Stein.
\newblock Reprogramming the pluripotent cell cycle: {R}estoration of an
  abbreviated {G}1 phase in human induced pluripotent stem (i{P}{S}) cells.
\newblock {\em J. Cell. Physiol.}, 226(5):1149--1156, 2011.

\bibitem{Wadkin19}
L.~E. {Wadkin}, S.~{Orozco-Fuentes}, I.~{Neganova}, S.~{Bojic}, A.~{Laude},
  M.~{Lako}, N.~G. {Parker}, and A.~{Shukurov}.
\newblock {Seeding hESCs to achieve optimal colony clonality}.
\newblock {\em Sci. Rep.}, 9:15299, 2019.

\bibitem{Strebinger19}
D.~Strebinger, C.~Deluz, E.~T. Friman, S.~Govindan, A.~B. Alber, and D.~M.
  Suter.
\newblock Endogenous fluctuations of oct4 and sox2 bias pluripotent cell fate
  decisions.
\newblock {\em Mol. Syst. Biol.}, 15(9):e9002, 2019.

\bibitem{Niwa00}
H.~Niwa, J.~Miyazaki, and A.~G. Smith.
\newblock Quantitative expression of {O}ct-3/4 defines differentiation,
  dedifferentiation or self-renewal of {E}{S} cells.
\newblock {\em Nat. Genet.}, 24(4):372--376, 2000.

\bibitem{Kopp08}
J.~L. Kopp, B.~D. Ormsbee, M.~Desler, and A.~Rizzino.
\newblock Small increases in the level of {S}ox2 trigger the differentiation of
  mouse embryonic stem cells.
\newblock {\em Stem cells}, 26(4):903--911, 2008.

\bibitem{Burykin14}
A.~Burykin, M.~D. Costa, L.~Citi, and A.~L. Goldberger.
\newblock Dynamical density delay maps: simple, new method for visualising the
  behaviour of complex systems.
\newblock {\em BMC Med. Inform. Decis. Mak.}, 14(1):6, 2014.

\bibitem{Fishman12}
M.~Fishman, F.~J. Jacono, S.~Park, R.~Jamasebi, A.~Thungtong, K.~A. Loparo, and
  T.~E. Dick.
\newblock A method for analyzing temporal patterns of variability of a time
  series from {P}oincar\'e plots.
\newblock {\em J. Appl. Physiol.}, 113(2):297--306, 2012.

\bibitem{Mandelbrot68}
B.~B. Mandelbrot and J.~W. Van~Ness.
\newblock Fractional brownian motions, fractional noises and applications.
\newblock {\em SIAM Rev}, 10(4):422--437, 1968.

\bibitem{Mielniczuk07}
J.~Mielniczuk and P.~Wojdy{\l}{\l}o.
\newblock Estimation of {H}urst exponent revisited.
\newblock {\em Comput. Stat. Data Anal.}, 51(9):4510--4525, 2007.

\bibitem{Zaret14}
K.~S. Zaret.
\newblock Genome reactivation after the silence in mitosis: recapitulating
  mechanisms of development?
\newblock {\em Dev. Cell}, 29(2):132--134, 2014.

\bibitem{Festuccia17}
N.~Festuccia, I.~Gonzalez, N.~Owens, and P.~Navarro.
\newblock Mitotic bookmarking in development and stem cells.
\newblock {\em Development}, 144(20):3633--3645, 2017.

\bibitem{Sveshnikov66}
A.~A. Sveshnikov, I.~N. Sneddon, and M.~Stark.
\newblock {\em Applied Methods of the Theory of Random Functions}.
\newblock ISSN. Elsevier Science, 1966.

\bibitem{Murray02}
J.~D. Murray.
\newblock {\em Mathematical Biology I. An Introduction}, volume~17 of {\em
  Interdisciplinary Applied Mathematics}.
\newblock Springer, New York, 3 edition, 2002.

\bibitem{Codling08}
E.~A. Codling, M.~J. Plank, and S.~Benhamou.
\newblock Random walk models in biology.
\newblock {\em J. R. Soc. Interface}, 6(25):813--834, 2008.

\bibitem{Li10}
L.~Li, B.~H. Wang, S.~Wang, L.~Moalim-Nour, K.~Mohib, D.~Lohnes, and L.~Wang.
\newblock Individual cell movement, asymmetric colony expansion, rho-associated
  kinase, and {E}-cadherin impact the clonogenicity of human embryonic stem
  cells.
\newblock {\em Biophys.}, 98:2442 -- 2451, 2010.

\bibitem{Wu14}
P.~Wu, A.~Giri, S.~X. Sun, and D.~Wirtz.
\newblock Three-dimensional cell migration does not follow a random walk.
\newblock {\em Proc. Natl. Acad. Sci. U.S.A.}, 111(11):3949--3954, 2014.

\bibitem{Wadkin17}
L.~E. Wadkin, L.~F. Elliot, I.~Neganova, N.~G. Parker, V.~Chichagova, G.~Swan,
  A.~Laude, M.~Lako, and A.~Shukurov.
\newblock Dynamics of single human embryonic stem cells and their pairs: a
  quantitative analysis.
\newblock {\em Sci. Rep.}, 7(1):1--12, 2017.

\bibitem{Wadkin18}
L.~E. Wadkin, S.~Orozco-Fuentes, I.~Neganova, G.~Swan, A.~Laude, M.~Lako,
  A.~Shukurov, and N.~G. Parker.
\newblock Correlated random walks of human embryonic stem cells in vitro.
\newblock {\em IOP Phys. Biol.}, 15(5):056006, 2018.

\bibitem{Chacondev18}
C.~A. Chac{\'o}n-Mart{\'\i}nez, J.~Koester, and S.~A. Wickstr{\"o}m.
\newblock Signaling in the stem cell niche: regulating cell fate, function and
  plasticity.
\newblock {\em Development}, 145(15), 2018.

\bibitem{Saha08}
S.~Saha, L.~Ji, J.~J. de~Pablo, and S.~P. Palecek.
\newblock Tgf$\beta$/activin/nodal pathway in inhibition of human embryonic
  stem cell differentiation by mechanical strain.
\newblock {\em Biophys. J.}, 94(10):4123 -- 4133, 2008.

\bibitem{Shahriyari13}
L.~Shahriyari and N.~L. Komarova.
\newblock Symmetric vs. asymmetric stem cell divisions: An adaptation against
  cancer?
\newblock {\em PLoS One}, 8(10):1--16, 10 2013.

\bibitem{Yamashita2010}
Y.~M. Yamashita, H.~Yuan, J.~Cheng, and A.~J. Hunt.
\newblock Polarity in stem cell division: asymmetric stem cell division in
  tissue homeostasis.
\newblock {\em CSH Perspect. Biol.}, 2(1):a001313--a001313, 2010.

\bibitem{Iskandar18}
R.~Iskandar.
\newblock A theoretical foundation for state-transition cohort models in health
  decision analysis.
\newblock {\em PLoS One}, 13(12):e0205543--e0205543, 2018.

\bibitem{LiDev19}
P.~Li and M.~B. Elowitz.
\newblock Communication codes in developmental signaling pathways.
\newblock {\em Development}, 146(12), 2019.

\bibitem{Chen14}
K.~G. Chen, B.~S. Mallon, K.~R. Johnson, R.~S. Hamilton, R.~D.G. McKay, and
  P.~G. Robey.
\newblock Developmental insights from early mammalian embryos and core
  signaling pathways that influence human pluripotent cell growth and
  differentiation.
\newblock {\em Stem Cell Research}, 12(3):610 --621, 2014.

\bibitem{Blauwkamp2012}
T.~A. Blauwkamp, S.~Nigam, R.~Ardehali, I.~L. Weissman, and R.~Nusse.
\newblock Endogenous {W}nt signalling in human embryonic stem cells generates
  an equilibrium of distinct lineage-specified progenitors.
\newblock {\em Nature Communications}, 3(1):1070, 2012.

\bibitem{Kurosaka08}
Satoshi Kurosaka and Anna Kashina.
\newblock Cell biology of embryonic migration.
\newblock {\em Embryo today: reviews}, 84:102 -- 122, 2008.

\bibitem{Fagotto3303}
Fran{\c c}ois Fagotto.
\newblock The cellular basis of tissue separation.
\newblock {\em Development}, 141(17):3303--3318, 2014.

\bibitem{KRENS2011189}
S.~F.~G. Krens and C-P. Heisenberg.
\newblock Chapter six - {C}ell {S}orting in {D}evelopment.
\newblock In Michel Labouesse, editor, {\em Forces and Tension in Development},
  volume~95 of {\em Current Topics in Developmental Biology}, pages 189 -- 213.
  Academic Press, 2011.

\bibitem{SirioQuantification}
Sirio Orozco-Fuentes, Irina Neganova, Laura~E. Wadkin, Andrew~W. Baggaley,
  Rafael~A. Barrio, Majlinda Lako, Anvar Shukurov, and Nicholas~G. Parker.
\newblock Quantification of the morphological characteristics of hesc colonies.
\newblock {\em Scientific Reports}, 9:17569, 2019.

\bibitem{Okamoto}
Kazuko Okamoto, Arno Germond, Hideaki Fujita, Chikara Furusawa, Yasushi Okada,
  , and Tomonobu~M. Watanabe.
\newblock Single cell analysis reveals a biophysical aspect of collective
  cell-state transition in embryonic stem cell differentiation.
\newblock {\em Scientific Reports}, 8:11965, 2018.

\bibitem{Purdom05}
E.~Purdom and S.~P. Holmes.
\newblock Error distribution for gene expression data.
\newblock {\em Stat. Appl. Genet. Mol. Biol.}, 4(1), 2005.

\bibitem{Holmes17}
W.~R. Holmes, N.~S. Reyes~de Mochel, Q.~Wang, H.~Du, T.~Peng, M.~Chiang,
  O.~Cinquin, K.~Cho, and Q.~Nie.
\newblock Gene expression noise enhances robust organization of the early
  mammalian blastocyst.
\newblock {\em PLoS Comput. Biol.}, 13(1):1--23, 01 2017.

\bibitem{Ghorbani18}
M.~Ghorbani, E.~A. Jonckheere, and P.~Bogdan.
\newblock Gene expression is not random: Scaling, long-range cross-dependence,
  and fractal characteristics of gene regulatory networks.
\newblock {\em Front. Physiol.}, 9:1446, 2018.

\bibitem{Liu15}
X.~Liu, B.~Wang, and L.~Xu.
\newblock Statistical analysis of {H}urst exponents of essential/nonessential
  genes in 33 bacterial genomes.
\newblock {\em PLoS One}, 10(6):1--9, 06 2015.

\bibitem{Bogdan14}
P.~Bogdan, B.~M. Deasy, B.~Gharaibeh, T.~Roehrs, and R.~Marculescu.
\newblock Heterogeneous structure of stem cells dynamics: statistical models
  and quantitative predictions.
\newblock {\em Sci. Rep.}, 4:4826, 2014.

\bibitem{Sieburg13}
H.~B. Sieburg, G.~Cattarossi, and C.~E. Muller-Sieburg.
\newblock Lifespan differences in hematopoietic stem cells are due to imperfect
  repair and unstable mean-reversion.
\newblock {\em PLoS Comput. Biol.}, 9(4):1--15, 04 2013.

\bibitem{Lacasa09}
L.~Lacasa, B.~Luque, J.~Luque, and J.~C. Nuno.
\newblock The visibility graph: A new method for estimating the {H}urst
  exponent of fractional {B}rownian motion.
\newblock {\em EPL}, 86(3):30001, 2009.

\bibitem{Barunik10}
J.~Barunik and L.~Kristoufek.
\newblock On {H}urst exponent estimation under heavy-tailed distributions.
\newblock {\em PHYSICA A}, 389(18):3844--3855, 2010.

\bibitem{Haghverdi15}
L.~Haghverdi, F.~Buettner, and F.~J. Theis.
\newblock {Diffusion maps for high-dimensional single-cell analysis of
  differentiation data}.
\newblock {\em Bioinformatics}, 31(18):2989--2998, 05 2015.

\bibitem{Akberdin18}
I.~R. Akberdin, N.~A. Omelyanchuk, S.~I. Fadeev, N.~E. Leskova, E.~A.
  Oschepkova, F.~V. Kazantsev, Y.~G. Matushkin, D.~A. Afonnikov, and N.~A.
  Kolchanov.
\newblock Pluripotency gene network dynamics: System views from parametric
  analysis.
\newblock {\em PLoS One}, 13(3):1--24, 03 2018.

\bibitem{Chickarmane06}
V.~Chickarmane, C.~Troein, U.~A Nuber, H.~M Sauro, and C.~Peterson.
\newblock Transcriptional dynamics of the embryonic stem cell switch.
\newblock {\em PLoS Comput. Biol.}, 2(9):1--13, 09 2006.

\bibitem{Likhoshvai07}
V.~Likhoshvai and A.~Ratushny.
\newblock Generalized {H}ill function method for modeling molecular processes.
\newblock {\em J. Bioinf. Comput. Biol.}, 05(02b):521--531, 2007.

\end{thebibliography}

\end{document}